\documentclass[12pt]{article}
\usepackage[]{axodraw}
\usepackage[]{epsfig}
\def\slash#1{\setbox0=\hbox{$#1$}#1\hskip-\wd0\hbox to\wd0{\hss\sl/\/\hss}}
\catcode`@=11
\newcount\@tempcntc
\def\@citex[#1]#2{\if@filesw\immediate\write\@auxout{\string\citation{#2}}\fi
  \@tempcnta\z@\@tempcntb\m@ne\def\@citea{}\@cite{\@for\@citeb:=#2\do
    {\@ifundefined
       {b@\@citeb}{\@citeo\@tempcntb\m@ne\@citea\def\@citea{,}{\bf ?}\@warning
       {Citation `\@citeb' on page \thepage \space undefined}}%
    {\setbox\z@\hbox{\global\@tempcntc0\csname b@\@citeb\endcsname\relax}%
     \ifnum\@tempcntc=\z@ \@citeo\@tempcntb\m@ne
       \@citea\def\@citea{,}\hbox{\csname b@\@citeb\endcsname}%
     \else
      \advance\@tempcntb\@ne
      \ifnum\@tempcntb=\@tempcntc
      \else\advance\@tempcntb\m@ne\@citeo
      \@tempcnta\@tempcntc\@tempcntb\@tempcntc\fi\fi}}\@citeo}{#1}}
\def\@citeo{\ifnum\@tempcnta>\@tempcntb\else\@citea\def\@citea{,}%
  \ifnum\@tempcnta=\@tempcntb\the\@tempcnta\else
   {\advance\@tempcnta\@ne\ifnum\@tempcnta=\@tempcntb \else \def\@citea{--}\fi
    \advance\@tempcnta\m@ne\the\@tempcnta\@citea\the\@tempcntb}\fi\fi}
\catcode`@=12

\def\theequation{\arabic{section}.\arabic{equation}}

\begin{document}

\begin{flushright}
  MZ-TH/97-28\\
  MPI/PhT/97-47\\
  UCSD/97-18\\
  hep-ph/9707517\\
  July 1997
\end{flushright}

\begin{center}
  {\LARGE {\bf Gauge-Independent Analysis of {\boldmath $K_L\to e\mu$}}}
                                                                 \\[0.4cm]
  {\LARGE {\bf in Left-Right Models}}\\[1.5cm]
  {\large Zoltan Gagyi-Palffy,}$^{\, a}$
  {\large Apostolos Pilaftsis}$^{\, b}$\footnote[2]{e-mail address: 
    pilaftsi@mppmu.mpg.de} 
  {\large and Karl Schilcher}$^{\, c}$\footnote[3]{Permanent address:
Institut f\"ur Physik, THEP, Johannes Gutenberg-Universit\"at,
    55099 Mainz,\\ Germany. E-mail address: 
    schilcher@dipmza.physik.uni-mainz.de}\\[0.4cm]
  $^a${\em Institut f\"ur Physik, THEP, Johannes Gutenberg-Universit\"at,
    55099 Mainz, Germany}\\[0.3cm]
  $^b${\em Max-Planck-Institut f\"ur Physik, F\"ohringer Ring 6, 80805
    Munich, Germany}\\[0.3cm]
  $^c${\em Department of Physics, Univ.\ of California, San Diego, 
       La Jolla, CA 92093-0319, USA}
\end{center}
\vskip0.7cm \centerline{\bf  ABSTRACT}   
The lepton-flavour-violating decay  $K_L\to e\mu$ is studied in detail
within    the  context   of      SU(2)$_R  \otimes$SU(2)$_L    \otimes
$U(1)$_{(B-L)}$  models,   which include   heavy  Majorana  neutrinos.
Particular  attention is paid to the  gauge independence of this decay
process   to one  loop.  In   analogy   with earlier  studies on   the
$K^0\bar{K}^0$ mixing, it is explicitly shown how restoration of gauge
invariance occurs in the  decay amplitude containing the box diagrams,
when the  relevant Higgs-dependent self-energy   and vertex graphs are
taken  into account in  the  on-shell skeleton renormalization scheme.
Based  on  the analytic  expressions  so  derived,  we find   that the
branching ratio  $B(K_L\to e\mu)$ can  be considerably enhanced due to
the presence of left-  and right-handed currents in  the loop, and can
reach values close  to or even   larger than the  present experimental
limit $3.3\times 10^{-11}$ in the manifest left-right symmetric model.
Constraints on  the parameter space of  typical left-right  models are
derived from the possible decay $K_L\to e\mu$ and a global analysis of
other low-energy data.

\newpage

\setcounter{equation}{0}
\section{Introduction}

Experiments involving kaons or kaon  decays have reached a high  level
of accuracy   and are therefore  very  sensitive to probe  new physics
beyond the minimal Standard   Model (SM).  For example, studying   the
small  mass difference $\Delta M_K$  between the long-lived kaon $K_L$
and the  short-lived one  $K_S$   is one such  measurement of  amazing
precision.  In the $K^0\bar{K}^0$ system, the observable $\Delta M_K /
M_K$  is  very  accurately  established at    the level of  $10^{-15}$
\cite{PDG}, in line with predictions in the SM \cite{CPreview}.  Other
experiments sensitive  to  new-physics   scenarios look for   possible
lepton-flavour-violating   decays  of the   type  $K_L\to  e\mu$.  The
experimental  upper bound on the  branching ratio  $B(K_L\to e\mu)$ is
very stringent \cite{PDG}, {\em i.e.},
\begin{equation}
\label{Bexp}
B_{\mbox{\scriptsize exp}}(K_L\to e\mu)\ <\ 3.3\times 10^{-11}\, ,
\end{equation}
at 90$\%$ confidence level.  Ongoing experiments at BNL, KEK and, more
recently, at DA$\Phi$NE are expected  to lower the  above bound by one
order  of magnitude.  Although   the decay $K_L\to  e\mu$  is strictly
forbidden in the SM due to the absolute conservation of the individual
leptonic   quantum numbers,   lepton-flavour  violation  occurs almost
inevitably in many of  its extensions. For  example, many studies have
been devoted  to analyze such a  decay in models based on technicolour
\cite{TCol},  supersymmetry  \cite{SUSY},      horizontal     symmetry
\cite{Hor},  extended      scalar   \cite{LQ}   and    gauge    sector
\cite{IL,CSL&TI,LSS,GPS}.   Thus, experiments on $K_L\to e\mu$ provide
a unique  opportunity either to  discover new physics or  to constrain
efficiently the parameter space of these theories.

In this paper, we shall present a careful analysis of $K_L\to e\mu$ in
left-right  models, which  are  realized by  the gauge  group SU(2)$_R
\otimes$SU(2)$_L$ $\otimes$U(1)$_{(B-L)}$  and include  heavy Majorana
neutrinos  \cite{MS}.  These theories  also  allow for scenarios  with
high Dirac mass terms of order of top  quark mass without invalidating
experimental limits on light neutrino masses.  In these scenarios, the
usual see-saw suppression relations   \cite{YAN} for  the  light-heavy
neutrino  mixings may be  avoided.   In particular, it  has been found
that the large Dirac masses give rise to a non-decoupling behaviour of
the heavy   neutrinos  in  loops   \cite{AP1,IP},  thereby   enhancing
significantly the decay   rate for  possible  lepton-flavour-violating
processes, such as $Z\to  e\tau$ \cite{KPS}  and  the size  of related
new-physics  observables at LEP1  \cite{Pepe,PRD}.  This feature makes
the left-right models   very attractive.  Apart from taking   properly
into consideration the non-decoupling   effect of heavy neutrinos,  we
shall also  analyze the  impact   of other   low-energy  data  on  the
theoretical  prediction  for $B(K_L \to  e\mu)$,  such as those coming
from  the  non-observation  of   the decay  $\mu\to   e\gamma$ and the
$K_LK_S$ mass difference.

Another  important issue, which is  to be discussed  in detail, is the
lack of gauge invariance,  when  only contributions from one-loop  box
diagrams to the decay $K_L \to e \mu$ are  considered.  Studies on the
$K^0\bar{K}^0$   system   revealed \cite{CSL&TI,WSH&AS,BLP}  that  box
diagrams by    themselves do not form a    gauge-invariant set of loop
graphs in the  left-right models.  In  addition to the box graphs, one
has to take into account  self-energy and vertex corrections involving
very  heavy  Higgs  bosons,  which  mediate flavour-changing   neutral
currents (FCNC).  Even if the FCNC Higgs bosons were taken to have all
infinite  masses, their effect would not   decouple from the loop.  In
the infinite  mass limit  for  the heavy  Higgs scalars, the  residual
terms  are also   gauge     dependent  and cancel   against    similar
gauge-dependent  terms coming from  the box graphs, leading eventually
to  a   gauge-invariant   result.    While  the    inclusion  of   the
gauge-dependent Higgs  self-energies  to the $K_LK_S$ mass  difference
leads to little  change in the value  predicted by the box graphs, the
situation for the decay $K_L\to e\mu$ is strikingly different. We find
that  the   residual   gauge-dependent  terms  originating  from   the
self-energy and vertex graphs may be comparable to or even larger than
the  box corrections.   As   a  consequence, the  constraints   on the
parameter space of   the left-right models will  now  become much more
severe than those  obtained in earlier   articles \cite{LSS,GPS}.  Our
calculation of the gauge-independent decay amplitude for $K_L\to e\mu$
at  one loop  and   the  analysis of  its   immediate phenomenological
consequence constitute novel aspects for  the left-right models, which
have not been studied in detail before.

The paper is organized as follows: In Section 2, we describe the basic
low-energy structure  of the left-right models  and review many useful
relations among  mixing parameters  and  neutrino  masses, as well  as
relations involving gauge-  and Higgs-boson masses.   In Section 3, we
calculate the  matrix element for the decay   $K_L\to e\mu$ and derive
mass limits on the FCNC Higgs  masses from the tree-level contribution
and other theoretical considerations.   All relevant Feynman rules and
one-loop analytic expressions pertaining to $K_L\to e\mu$ are given in
Appendices A  and B, respectively.  In  Section  4, we show  how gauge
independence gets restored, when Higgs-mediated self-energy and vertex
graphs are  taken into account in addition  to the box diagrams within
the well-defined on-shell skeleton renormalization scheme.  A detailed
discussion on the dependence  of $B(K_L\to e\mu)$ on various kinematic
parameters  follow in Section  5.  Numerical estimates for the  actual
size  of $B(K_L\to e\mu)$   are  given  within  manifest  as   well as
non-manifest left-right symmetric models. This  enables one to  deduce
combined constraints   on    the parameters  of   these   models.  Our
conclusions are summarized in Section 6.

\setcounter{equation}{0}
\section{SU(2)$_R\otimes$SU(2)$_L\otimes$U(1)$_{(B-L)}$ model}

The  idea of having  spontaneous breakdown of  both gauge and discrete
symmetries can  naturally be realized in  left-right theories based on
the    gauge  group  SU(2)$_R \otimes$SU(2)$_L  \otimes$U(1)$_{(B-L)}$
\cite{PS,FM,MS}, where $B$  and $L$ denote  baryon and lepton numbers,
respectively.  Such a realization may  arise from the low-energy limit
of SO(10) grand   unified theories.  Specifically,   one of the   most
appealing breaking patterns of SO(10) down to the SM is the following:
\begin{eqnarray}
{\rm SO}(10) &\to& {\rm SU}(4)_{\rm PS}\otimes {\rm SU(2)}_R\otimes 
{\rm SU(2)}_L \ \to\  
{\rm SU}(3)_c \otimes {\rm SU}(2)_R\otimes {\rm SU}(2)_L \otimes 
                                          {\rm U}(1)_{(B-L)} \nonumber\\
       &\to& {\rm SU}(3)_c \otimes {\rm SU}(2)_L \otimes {\rm U}(1)_Y.
\end{eqnarray}
Among the several  Higgs scalar representations  which can lead to the
SM,   we  shall adopt  the   field   content of the  left-right  model
introduced    first  in   \cite{MS}, which  includes    heavy Majorana
neutrinos.

We will now describe the  basic low-energy structure of the left-right
models.  In a  three-generation left-right model, the  left-handed and
right-handed chiral fields are  grouped in separate  weak iso-doublets
as follows:
\begin{eqnarray}
\label{Qnumbers}
L_L' = \left( \begin{array}{c} \nu'_l \\ l' \end{array} \right)_L\ :\ 
(0,1/2,-1) &&\qquad
L_R' = \left( \begin{array}{c} \nu'_l \\ l' \end{array} \right)_R\ :\ 
(1/2,0,-1),\\
Q_L' = \left( \begin{array}{c} u' \\ d' \end{array} \right)_L\ :\ 
 (0,1/2,1/3) &&\qquad
Q_R' = \left( \begin{array}{c} u' \\ d' \end{array} \right)_R\ :\ 
(1/2,0,1/3),
\end{eqnarray}
where the  weak   eigenstates  are  denoted   by  a prime.    In  Eq.\
(\ref{Qnumbers}), we have  also  indicated the assignment of  quantum
numbers         to  fermions     under          the    gauge     group
SU(2)$_R\otimes$SU(2)$_L\otimes$U(1)$_{(B-L)}$.  Evidently, the  above
left-right symmetric extension  of  the SM gauge  sector  requires the
presence of right-handed neutrinos.

The Higgs fields introduced  into the model  to break left-right gauge
symmetry down  to  $U(1)_{em}$ \cite{MS} are:
\begin{equation}
\Phi\ =\ \left( \begin{array}{cc} \phi^0_1 & \phi^+_1 \\
                          \phi^-_2 & \phi^0_2 \end{array} \right),\quad
\Delta_L\ =\ \left( \begin{array}{cc} \delta_L^+/\sqrt{2} & \delta_L^{++} \\
                               \delta_L^0 & -\delta_L^+/\sqrt{2}
\end{array} \right),\quad
\Delta_R\ =\ \left( \begin{array}{cc} \delta_R^+/\sqrt{2} & \delta_R^{++} \\
                               \delta_R^0 & -\delta_R^+/\sqrt{2}
\end{array} \right), 
\end{equation}
where $\Phi$ is a  $(1/2^*,1/2,0)$ Higgs bidoublet, and $\Delta_L$ and
$\Delta_R$ are  two  complex   Higgs  triplets with  quantum   numbers
$(0,1,2)$    and   $(1,0,2)$,    respectively.   To   avoid  excessive
complication, we will  assume  that only  $\phi^0_1$ and  $\delta^0_R$
acquire vacuum expectation values (VEV's): $\langle \phi^0_1 \rangle =
\kappa_1/\sqrt{2}$  and  $\langle   \delta^0_R\rangle = v_R/\sqrt{2}$.
Such a     scheme   is also  favoured  for    phenomenological reasons
\cite{GGMKO}.  At the tree level \cite{GSenj,CSL&TI}, the vanishing of
the VEV's for  the neutral fields $\phi^0_2$  and $\delta_L^0$ can  be
achieved by imposing invariance   of  the left-right  symmetric  Higgs
potential under the discrete symmetry D: $\Phi\to i\Phi$, $\Delta_L\to
\Delta_L$   and  $\Delta_R\to -  \Delta_R$.   As  has been observed in
\cite{CM}, however,  the   D symmetry may suppress   unnaturally large
FCNC's but  leads to zero masses  for the down  quarks and the charged
leptons.   Even though   the   latter  may be   viewed as   a   viable
approximation for $K_L\to e\mu$,  the  presence of non-zero  $m_{d_i}$
and  $m_{l_i}$ would  require infinite renormalization   of the VEV of
$\delta_L^0$.  To  overcome  this problem,  the  authors  in \cite{CM}
suggested  that stability  of such a  model  under high order  quantum
corrections can  naturally  be  achieved within  a   D-parity breaking
scenario. Since we shall treat charged leptons and down-type quarks as
being strictly  massless at the quantum  level, the above problem does
not occur in our calculations.

The left-right scenario of our interest is  further described in Ref.\
\cite{GGMKO}, under case (d).  The fact that $\langle\delta^0_L\rangle
= \langle \phi^0_2 \rangle =  0$ in this case  implies the absence  of
mixing between the  charged gauge bosons  $W^+_L$ and $W^+_R$.   Thus,
$W^+_L$  may be identified as  the  known $W^+$  boson with mass $M_W$
observed experimentally,   whereas $W^+_R$ is   an  extra gauge  boson
having a  mass  $M_R$ larger  than  at  least  0.5  TeV \cite{PDG}. In
addition, the minimal  left-right  model predicts two massive  neutral
gauge  bosons, $Z_L$ and  $Z_R$, which have a non-vanishing tree-level
mixing   of order $\kappa_1^2   / v_R^2  \sim 10^{-2}$.  Nevertheless,
$Z_{L,R}$-mediated interactions are not of  much relevance here, since
they do not directly enter our discussion of $K_L\to e\mu$.

The scalar spectrum of the left-right model consists  of 20 degrees of
freedom, of which six  degrees become  the longitudinal components  of
the gauge  bosons: $W^\pm_L$,  $W^\pm_R$, $Z_L$  and $Z_R$.  From  the
remaining 14 physical  states, only the  charged scalars  $h^\pm$, the
CP-even Higgs boson $\phi^r_2 = \Re e\phi^0_2/\sqrt{2}$ and the CP-odd
Higgs scalar   $\phi^i_2  =  \Im m\phi^0_2/\sqrt{2}$  are   of  direct
relevance to   the  decay  $K_L\to   e\mu$.  The  field  $h^+$  is the
orthogonal  component  of  the  right-handed  would-be Goldstone boson
$G^+_R$.  The fields $h^+$ and  $G^+_R$ are not pure  mass-eigenstates
but admixtures of $\phi^+_1$ and $\delta^+_R$, {\em viz.}
\begin{equation}
\label{h+GR}
\left( \begin{array}{c} h^+ \\ G^+_R \end{array} \right)\ =\ 
\left( \begin{array}{cc} c_\beta & s_\beta \\ -s_\beta & c_\beta 
\end{array} \right) \left( \begin{array}{c} \phi^+_1 \\ \delta^+_R 
\end{array} \right)\, ,\qquad s_\beta\ =\ \sqrt{1-c^2_\beta}\ =\ 
\frac{M_W}{M_R}\ .
\end{equation}
As we will see in  Section 3, the scalars  $\phi^{r,i}_2$ must be very
heavy, {\em i.e.}, heavier than  10 TeV, in order  to avoid large FCNC
contributions   to the $K_LK_S$  mass  difference.  Furthermore, after
diagonalization  of    the relevant  mass   matrices,    $h^\pm$   and
$\phi^{r,i}_2$ are found to be nearly degenerate \cite{GGMKO}. In
fact, one obtains
\begin{equation} 
  \label{Masrel}
M^2_{\phi^r_2}\ =\ M^2_{\phi^i_2}\, +\, {\cal O}(M^2_W)\ =\ M^2_h\, +\,
{\cal O}(M^2_W)
\end{equation}
Therefore, it is reasonable to  assume that the Higgs scalars $h^\pm$,
$\phi^r_2$ and $\phi^i_2$ are {\em exactly} degenerate.

The left-right model  admits the presence of $B-L$-violating operators
in the  Yukawa sector,  which  are introduced  by the   triplet fields
$\Delta_{L,R}$ in the following way:
\begin{equation}
{\cal L}^{B-L}_{\rm int}\ =\ - \frac{\sqrt{2} m_{M_{ij}}}{2v_R}\Big(h_{ij}
\bar{L}'^C_{L_i}\, \varepsilon_{ij} \Delta_L L_{L_j}'\ +\ 
\bar{L}'^C_{R_i}\, \varepsilon_{ij} \Delta_R L_{R_j}'\Big)\quad +\quad 
\mbox{H.c.},
\end{equation}
where $\varepsilon_{ij}$  is  the usual Levi-Civita  tensor, $m_{Mij}$
are Majorana mass terms  and $h_{ij}$ are  Yukawa couplings, which are
much smaller than  unity  owing to  phenomenological  constraints from
muon and   $\tau$  decays  \cite{GGMKO}.  After  spontaneous  symmetry
breaking, the Lagrangian describing the  neutrino mass matrix is given
by
\begin{eqnarray}
\label{LnuMass}
{\cal L}^\nu_{\rm mass} &=& -\frac{1}{2}\, 
\left( \bar{\nu}_L', \ \bar{\nu}'^{C}_R \right)\, M^\nu
\left( \begin{array}{c} \nu'^C_L \\ \nu'_R \end{array} \right)\ +\ 
       {\rm H.c.} \nonumber\\
&=& -\frac{1}{2}\ \bar{n}_L \widehat{M}^\nu n_R\ +\ {\rm H.c.}, 
\end{eqnarray}
where the $6\times 6$-dimensional  matrix  $M^\nu$ takes on  the known
see-saw-type form, {\em i.e.},
\begin{equation}
\label{Mnu}
M^\nu\ =\ \left( \begin{array}{cc} 0 & m_D\\ m_D^T & m_M
\end{array}\right)\, .
\end{equation} 
The   Dirac mass matrix  $m_D$   is a  $3\times 3$-dimensional  matrix
originating from the breaking   of the SU(2)$_L$ sector.   In general,
the matrix $M^\nu$ is always  diagonalizable by a $6\times 6$  unitary
matrix $U^\nu$ through
\begin{equation}
U^{\nu T}\, M^\nu\, U^\nu\ =\ \widehat{M}^\nu\, .
\end{equation}
After diagonalization, one gets six physical Majorana neutrinos $n_i$.
Three of  the six mass-eigenstates are  assigned to the ordinary light
neutrinos, $\nu_e$, $\nu_\mu$   and $\nu_\tau$,  while the   remaining
three states, $N_1$, $N_2$ and $N_3$, describe new Majorana neutrinos,
which must be heavier than about 100  GeV for phenomenological reasons
\cite{ZPC}.

By analogy,   the $3\times  3$  charged-lepton mass  matrix  $M^l$ can
always be diagonalized through the bi-unitary transformation
\begin{equation}
V_L\, M^l\, V_R^\dagger\ =\ \widehat{M}^l\, ,
\end{equation}
where $V_L$ and $V_R$ are unitary matrices that relate the weak fields
$l'_L$  and $l'_R$  to    the  mass  eigenstates  $l_L$  and    $l_R$,
respectively. Adopting the conventions of Ref.\ \cite{ZPC}, we can now
write the charged-current  interaction of the Majorana neutrinos $n_i$
and  charged leptons $l_i$ with  the left-handed gauge boson $W^\pm_L\
(\equiv W^\pm)$.  The charged-current Lagrangian is given by
\begin{eqnarray}
\label{WLlnu}
{\cal L}^{W_L}_{\rm int} &=& -\frac{g_w}{\sqrt{2}} W_L^{-\mu}\, B^L_{li}\ 
\bar{l}\gamma_\mu \mbox{P}_L n_i\quad +\quad \mbox{H.c.},
\end{eqnarray}
where $g_w$ is the weak coupling constant, $c^2_w = 1  - s^2_w = M^2_W
/ M^2_Z$,   $\mbox{P}_{L(R)} = [1-(+)\gamma_5  ]/2$   and $B^L$ is the
  $3\times 6$ matrix
\begin{equation}
\label{BL}
B^L_{lj}\ =\ \sum\limits_{k=1}^{3} V_{L_{lk}} U^{\nu\ast}_{kj}\, ,
\end{equation}
which describes the mixing of the left-handed charged leptons with the
neutrinos.  Furthermore, the    $Z_L$ boson couples  to  the  Majorana
neutrinos through the $6\times 6$ mixing matrix
\begin{equation} 
\label{CL}
C^L_{ij}\ =\ \sum\limits_{k=1}^{3}\ U^\nu_{ki}U^{\nu\ast}_{kj}. 
\end{equation} 
In  an  analogous  way,  the mixing   matrices  $B^R_{li}$ and
$C^R_{ij}$,
\begin{equation}
B^R_{lj}\ = \sum\limits_{k=4}^{6} V_{R_{lk}} U^{\nu}_{kj}\quad
\mbox{and}\quad
C^R_{ij}\ =\ \sum\limits_{k=4}^{6}\ U^{\nu\ast}_{ki}U^\nu_{kj},
\label{BR}
\end{equation} 
mediate the interactions   of the $W^\pm_R$ and  $Z_R$   bosons to the
charged leptons and Majorana neutrinos.  In the left-right models, the
quark sector  is   modified  as well,  namely,   there  are additional
couplings of the fermionic fields to the gauge and extra Higgs bosons.
All Feynman  rules that are  necessary for  the calculation of $K_L\to
e\mu$ are relegated to Appendix A.

The  left-handed flavour-mixing  matrices $B^L$  and  $C^L$, which are
identical to the respective mixing matrices $B$ and $C$ of the SM with
right-handed neutrinos, satisfy  a  number of identities that   may be
found in \cite{ZPC,IP}.  In addition, the right-handed mixing matrices
$B^R$ and $C^R$ obey the relations \cite{PRD}:
\begin{equation}
\label{idR}
\sum\limits_{i=1}^{6} B^L_{l_1i}B^R_{l_2i}\ =\  0,\qquad
\sum\limits_{l=1}^{3} B^{R\ast}_{li}B^R_{lj}\ =\  C^R_{ij},\qquad
C^{L\ast}_{ij}\ +\ C^R_{ij}\ = \ \delta_{ij}. 
\end{equation} 
These identities  is a direct  consequence  of  the unitarity  of  the
theory, assuring    its renormalizability.  To   simplify  further the
analysis of $K_L\to e\mu$, we shall assume  that only the electron and
muon families mix    effectively,  whereas the  $\tau$  family   stays
unmixed. In such an effective two-generation-mixing model, one obtains
\cite{AP1,IP}
\begin{equation}
\label{BLN2mix}
B^L_{lN_1}\ =\ \frac{\rho^{1/4} s^{\nu_l}_L}{\sqrt{1+\rho^{1/2}}}\ , \qquad
B^L_{lN_2}\ =\ \frac{i s^{\nu_l}_L}{\sqrt{1+\rho^{1/2}}}\ ,
\end{equation}
where the ratio  $\rho=m^2_{N_2}/m^2_{N_1}$ is always understood to be
greater or equal  than unity and $s^{\nu_l}_L$  (with $l = e,\mu$) are
the light-heavy neutrino mixings introduced in \cite{LL}, {\em i.e.},
\begin{equation}
\label{snul}
(s^{\nu_l}_L)^2 \ \equiv\  \sum\limits_{i=1}^{2} |B^L_{lN_i}|^2
\ \simeq\ \left( m_D^\dagger \frac{1}{m_M^2} m_D \right)_{ll}.   
\end{equation}
In addition, for the mixings $C^L_{N_iN_j}$, one finds
\begin{eqnarray}
\label{CL2mix}
C^L_{N_1N_1} &=& \rho^{1/2}\, C^L_{N_2N_2}\ =\
\frac{\rho^{1/2}}{1+\rho^{1/2}}\ [(s^{\nu_e}_L)^2
+(s^{\nu_\mu}_L)^2]\, , \nonumber\\
C^L_{N_1N_2}&=& -C^L_{N_2N_1}\ =\ i\rho^{1/4}\, C^L_{N_2N_2}\, .
\end{eqnarray} 
Finally,    with   the  help    of    Eq.~(\ref{idR})  together   with
Eq.~(\ref{BR}),  we can find  the    mixings  $B^R_{lN_i}$ up to    an
arbitrary   Cabbibo-type angle   $\theta_R$ \cite{PRD}.    The mixings
$B^R_{lN_i}$ are given by
\begin{eqnarray}
\label{BRmix}
B^R_{l_1N_1} &=& \cos\theta_R\sqrt{1-C^L_{N_1N_1}}\ ,\qquad
B^R_{l_2N_1} \ =\ -\sin\theta_R\sqrt{1-C^L_{N_1N_1}}\ ,\nonumber\\
B^R_{l_1N_2} &=& -\cos\theta_R\frac{C^L_{N_2N_1}}{\sqrt{1-C^L_{N_1N_1}}}
\, -\, i\sin\theta_R\, \Big[\, \frac{(1-C^L_{N_1N_1})(1-C^L_{N_2N_2})-
|C^L_{N_1N_2}|^2 }{1-C^L_{N_1N_1}}\, \Big]^{1/2},
\nonumber\\
B^R_{l_2N_2} &=& \sin\theta_R\frac{C^L_{N_2N_1}}{\sqrt{1-C^L_{N_1N_1}}}
\, -\, i\cos\theta_R\, \Big[\, \frac{(1-C^L_{N_1N_1})(1-C^L_{N_2N_2})-
|C^L_{N_1N_2}|^2 }{1-C^L_{N_1N_1}}\, \Big]^{1/2}. \ \ \
\end{eqnarray}
Counting now the number of independent kinematic parameters present in
the  leptonic sector of our  two-generation left-right  model, we find
five free quantities: the masses of the  two heavy Majorana neutrinos,
$m_{N_1}$  and $m_{N_2}$ [or equivalently   $m_{N_1}$ and $\rho$], the
two     light-heavy   neutrino    mixings,    $(s^{\nu_e}_L)^2$    and
$(s^{\nu_\mu}_L)^2$, which are severely constrained by low-energy data
as we will see later on in this section, and the angle $\theta_R$.

It  is now interesting to   remark  that $M^\nu$  in Eq.\  (\ref{Mnu})
reduces to   the traditional  seesaw scheme  \cite{YAN}   in the limit
$m_M\gg m_D$.   In the see-saw mass  pattern, the light-heavy neutrino
mixings  $s^{\nu_l}_L$  are very  suppressed, since  they are governed
from the  relation $s^{\nu_l}_L \sim \sqrt{m_{\nu_l}/m_N}$.  The usual
seesaw  models  predict very  small rates for lepton-flavour-violating
decays, such  as $K_L\to e\mu$,  $\mu\to eee$,  {\em etc.},  which are
beyond  the realm  of  detection in any  foreseeable  experiment.  The
situation may  change drastically if a  non-trivial mixing between two
families \cite{WW}  is introduced.  To give  an example,  consider the
following Dirac and Majorana sub-matrices in Eq.\ (\ref{Mnu}):
\begin{equation}
\label{example}
m_D\ =\ \left( \begin{array}{cc} 0 & a \\ 0 & b \end{array} \right)\, , 
\qquad
m_M\ =\ \left( \begin{array}{cc} 0 & A \\ A & \mu \end{array} \right)\, ,
\end{equation}
where  $a$, $b$, $A$  and $\mu$ are  arbitrary  complex numbers. It is
then easy to verify that the rank of the $4\times 4$ matrix $M^\nu$ is
two.  This implies that two   eigenstates are exactly massless at  the
tree level,  whereas the  other two are  massive.  The  scheme in Eq.\ 
(\ref{example}) may also be derived from an horizontal symmetry, which
is broken softly by a lepton-number-violating operator proportional to
$\mu$.  In fact,  if $\mu$ arises  from the spontaneous breakdown of a
global symmetry, then it should be $\mu \ll A$ in order to avoid large
Majoron couplings to electron and $u$, $d$ quarks, as is required from
astrophysical considerations \cite{PRD2}.  In this scenario, the light
neutrinos  acquire  masses radiatively in agreement  with experimental
upper bounds  \cite{ZPC}.  Most importantly, the mixings $s^{\nu_l}_L$
now scale as $s^{\nu_l}_L \sim a/A,\ b/A$ and therefore are large.  In
particular,  we should note that the  derivation  of the mixings $B^L$
and $B^R$ in   Eqs.\  (\ref{BLN2mix}) and (\ref{BRmix})  are  based on
non-seesaw scenarios of  the   generic form  (\ref{example}).    These
mixings can  only be constrained from a  global analysis of low-energy
data.

There are many experimental  data that can  be  used to constrain  the
light-heavy neutrino mixings \cite{LL}.  Here,  we shall however focus
only on those limits that  involve the $e$  and $\mu$ families and are
relevant  for the decay $K_L\to e\mu$.   We shall also assume that the
two heavy Majorana neutrinos $N_1$ and  $N_2$ are almost degenerate in
mass,  namely,  $m_{N_1} =  m_{N_2}  =  m_N$.  Thus,   charged current
universality in  $\pi$ decays \cite{KP}   and   in the decay   $\mu\to
e\nu\nu$ \cite{PDG} lead to the limits:
\begin{equation}
(s^{\nu_e}_L)^2,\  \  (s^{\nu_\mu}_L)^2 \ <\ 0.010\, .
\end{equation}
Furthermore, in  the large $m_N$ limit,  heavy Majorana neutrinos give
rise to $\mu\to e\gamma$ decays with branching ratio
\begin{equation}
\label{muegamma}
B(\mu \to e\gamma) \ =\ \frac{3\alpha_{em}}{8\pi}\, 
(s^{\nu_e}_L)^2\, (s^{\nu_\mu}_L)^2\, .
\end{equation}
On the experimental side, $B_{\rm exp} (\mu \to e\gamma) \le 4.9\times
10^{-11}$, which yields the tight constraint:
\begin{equation}
\label{mix}
(s^{\nu_e}_L)^2 (s^{\nu_\mu}_L)^2\ < \ 5.6\times 10^{-8}\, .
\end{equation}       
For the  model at hand,  the absence  of $\mu-e$ conversion  events in
nuclei \cite{JDV}   may lead  to more stringent    bounds due  to  the
non-decoupling behaviour of    the  heavy neutrinos  in  the   $Ze\mu$
coupling \cite{KPS}. A first estimate gives
\begin{equation}
\label{mueconv}
\frac{1}{2}\, s^{\nu_e}_Ls^{\nu_\mu}_L\, 
[(s^{\nu_e}_L)^2 + (s^{\nu_\mu}_L)^2]\ \stackrel{\displaystyle
  <}{\sim}\ 10^{-6}\times\, \frac{M^2_W}{m^2_N}\, .
\end{equation}
If the heavy neutrinos have TeV mass, both constraints (\ref{mix}) and
(\ref{mueconv})   are then  comparable.  For  very  heavy neutrinos of
order 10 TeV, the bound due to $\mu-e$ conversion is more restrictive.
Of course, one can completely avoid the latter constraint by assuming,
for  example, that  the muon  number is  practically  conserved by the
left-handed interactions,  which  merely means that  $b\ll a$  in Eq.\
(\ref{example}).  Even  if one  takes $s^{\nu_\mu}_L=0$, $K_L\to e\mu$
can still proceed via  the exchange of  $W^+_L$ and $W^+_R$  bosons in
the loop.  Finally, we find that  the decay $\mu  \to eee$ provides in
general a weaker constraint than that given in Eq.\ (\ref{mueconv}).

In the  see-saw type matrix (\ref{Mnu}), the  Dirac and  Majorana mass
terms,  $m_D$  and    $m_M$, cannot  be   arbitrarily  large   but are
constrained from above  by   triviality  and  perturbative   unitarity
bounds.  Renormalization-group-triviality   bounds, which   are mainly
controlled  by large contributions  to the quartic scalar couplings in
the Higgs potential, usually lead to the limit: $m_D < 2M_W/g_w\approx
300$  GeV.  However, the  coexistence of large Majorana masses $m_M\gg
m_D$  in the loop  of  the  Higgs potential  will screen  the one-loop
contributions    of the  heavy    neutrinos  to  the Higgs   potential
considerably.   In fact, the low-energy  data  mentioned above require
$m_D/m_M < 10^{-1}$.  Therefore, a more reliable constraint arises due
to perturbative unitarity, where the inequality $\Gamma (N\to lW^+_L,\
\nu Z_L) < m_N/2$ is imposed.  This inequality amounts to the bound
\begin{equation}
\label{mDbound}
m_D\ \stackrel{\displaystyle <}{\sim}\ 10\,  M_W\, .
\end{equation}
Similarly, the Majorana    mass   $m_M$ may be  constrained    from an
analogous  inequality $\Gamma  (N\to lW^+_R,\ \nu  Z_R)< m_N/2$, which
gives the bound
\begin{equation}
\label{mMbound}
m_M\ \stackrel{\displaystyle <}{\sim}\ 10\, M_R\, .
\end{equation}
The limits  derived in  Eqs.\ (\ref{mDbound}) and  (\ref{mMbound}) are
used throughout our analysis.

\setcounter{equation}{0}
\section{Matrix element of $K_L\to e\mu$}

In the left-right models, the lepton-flavour violating decay $K_L\to e
\mu$  can occur  at the  tree  level via  the exchange   of FCNC Higgs
scalars, as shown   in Fig.\ 1.   However,  for very heavy  FCNC Higgs
masses, the tree-level contribution gets rather  suppressed due to the
intermediate Higgs-boson  propagator, while  one-loop graphs shown  in
Figs.\ 2--4 can, in principle,  become very significant, since they do
not depend explicitly on  the FCNC Higgs  masses. Therefore,  we shall
include  in our study all  the relevant one-loop diagrams contributing
to $K_L \to e \mu$.

The  one-loop matrix element  of   the decay  $K_L  \to e\mu$  has the
general form
\begin{equation}
\label{Tmatrix}
{\cal T}(\bar{K}^0\to e^+\mu^- )\ =\ \frac{1}{(4\pi )^2}
\left( \frac{g_w}{\sqrt{2}}\right) ^4  \frac{1}{M_{W}^2}\,
\langle 0| \bar{d} \gamma_{\kappa} \mbox{P}_L s |
\bar{K} ^0\rangle\, \bar{u}_{\mu} \gamma^{\kappa} \mbox{P}_L
v_e \, ( {\cal A}\, +\, \eta {\cal B} )\, ,
\end{equation}
where  the reduced amplitudes  ${\cal A}$ and ${\cal  B}$ are given in
Appendix  B.  The  parameter $\eta$  in   Eq.\ (\ref{Tmatrix})  is  an
enhancement factor originating from chirality-flipping operators which
enter in the kaon-to-vacuum matrix element. The factor $\eta$ is given
by
\begin{eqnarray}
\label{eta}
\eta &=&
\frac{\langle 0 | \bar{d}\gamma^{\sigma}\gamma^{\kappa}
\mbox{P}_R\, s | \bar{K}^{0}\rangle\, \bar{u}_{\mu}\gamma_{\sigma}
\gamma_{\kappa}\mbox{P}_L v_{e}}{  
\langle 0 | \bar{d}\gamma^{\alpha}\mbox{P}_L\, s| 
\bar{K}^{0}\rangle\, \bar{u}_{\mu}
\gamma_{\alpha} \mbox{P}_L v_{e}}\nonumber\\
& \simeq & \frac{4M_K^2 }{(m_s+m_d)m_{\mu}}\simeq 50\, ,
\end{eqnarray}
where $M_K$ is the   mass of $K^0$,   and $m_d$ and $m_s$  denote  the
current  $d$- and $s$-quark masses,  respectively \cite{Gas/Leut}.  In
the  derivation of the  parameter $\eta$ in  Eq.\ (\ref{eta}), we have
taken into account the PCAC hypothesis and the fact that
\begin{equation}
\label{hadrMatrix}
\langle 0 | \bar{d}\gamma_{\mu}\gamma_{\nu}
\mbox{P}_R\, s | \bar{K}^{0}(p)\rangle\ =\  g_{\mu\nu}\, 
\langle 0|\bar{d}\gamma_5 s|\bar{K}^0(p)\rangle\, +\, \frac{1}{2} \,
\langle 0|\bar{d}\, [\gamma_\mu,\gamma_\nu]\, \gamma_5
s|\bar{K}^0(p)\rangle\, .
\end{equation}
The  second term   on the RHS  of  Eq.\  (\ref{hadrMatrix})  must be a
Lorentz  tensor antisymmetric  in   $p^\mu p^\nu$ and   hence vanishes
identically.  Applying the   PCAC  hypothesis to the  hadronic  matrix
element
\begin{equation}
  \label{fkaon1}
\langle 0|\bar{d}(x)\gamma_\mu\gamma_5 s(x)|\bar{K}^0 (p)\rangle\ =\
ip_\mu\, f_K e^{-ipx}\, ,
\end{equation}
where $f_K$ is  the kaon decay constant,  we can  calculate the matrix
element
\begin{equation}
\label{fkaon2}
 \langle 0|\bar{d}(x)\gamma_5 s(x)|\bar{K}^0 (p) \rangle\ =\
 -i\, \frac{M^2_K f_K}{m_d+m_s}\ e^{-ipx}\, . 
\end{equation}
Using the relations (\ref{hadrMatrix}), (\ref{fkaon1}), (\ref{fkaon2})
and little algebra,  the second equality  in Eq.\ (\ref{eta})  follows
easily.

Substituting  Eq.\  (\ref{fkaon1}) into Eq.\ (\ref{Tmatrix}),  we find
the branching ratio
\begin{equation}
  \label{BRKL}
B(K_L\to e\mu )\ =\ 4.1\times 10^{-4}\ | {\cal A}\, +\, \eta {\cal B}|^2\, ,
\end{equation}
where the charge of the leptons in the final state is not tagged.  One
arrives at the same result for $B(K_L \to e\mu)$ in Eq.\ (\ref{BRKL}),
by using relations   based  on isospin  invariance  between  the decay
amplitudes of $\bar{K}^0\to \mu^-e^+$ and $K^-\to \mu^-\nu_\alpha$.

\begin{center}
\begin{picture}(120,100)(0,0)
\SetWidth{0.8}
\ArrowLine(0,20)(30,50)\ArrowLine(30,50)(0,80)
\DashLine(30,50)(90,50){5}\ArrowLine(90,50)(120,20)\ArrowLine(120,80)(90,50)
\Text(0,90)[l]{$\bar{d}$}\Text(0,10)[l]{$s$}\Text(120,90)[r]{$e^+$}
\Text(120,10)[r]{$\mu^-$} \Text(60,60)[]{$\phi^{r,i}_2$}

\end{picture}\\[0.7cm]
{\small {\bf Fig.\ 1:}} {\small Tree-level graph contributing to
  $K_L\to e\mu$}
\end{center}

The tree-level $\phi^{r,i}_2$-exchange shown  in Fig.\ 1 gives rise to
a contribution to ${\cal T}(\bar{K}^0\to e\mu)$, which is proportional
to the reduced amplitude
\begin{eqnarray}
\label{FCNCtree} 
{\cal B}_{\mbox{{\scriptsize tree}}}& =& 
\frac{\pi}{\alpha_w}\, \sum_{i=c,t}\, \sum_{\alpha =1}^{6}
( \lambda_i\lambda_\alpha )^{1/2} \Big[\, \frac{1}{\lambda_R}\, 
\Big( 
V^{L*}_{id}V^R_{is}B^{L}_{\mu\alpha}B^{R*}_{e\alpha}\, +\, V^{L*}_{id}V^R_{is}
B^{R}_{\mu\alpha}B^{L*}_{e\alpha}\, +\ (L\leftrightarrow R)\Big)\nonumber\\
&&+\, \frac{1}{\lambda_I}\, \Big( V^{L*}_{id}V^R_{is}
B^{L}_{\mu\alpha}B^{R*}_{e\alpha} - V^{L*}_{id}V^R_{is}
B^{R}_{\mu\alpha}B^{L*}_{e\alpha}\, +\ (L\leftrightarrow R)\Big)\Big]\, ,
\end{eqnarray}
whereas  ${\cal    A}_{\mbox{{\scriptsize    tree}}}=0$.  In      Eq.\ 
(\ref{FCNCtree}), we  have defined $\lambda_{c,t}  = m^2_{c,t}/M^2_W$,
$\lambda_\alpha = m^2_{n_\alpha} / M^2_W$, $\lambda_R = M^2_{\phi^r_2}
/ M^2_W$ and $\lambda_I = M^2_{\phi^i_2} / M^2_W$.

\begin{center}
\begin{picture}(400,100)(0,0)
\SetWidth{0.8}
\ArrowLine(0,30)(30,30)\ArrowLine(30,30)(30,70)\ArrowLine(30,70)(0,70)
\ArrowLine(70,30)(100,30)\ArrowLine(70,70)(70,30) \ArrowLine(100,70)(70,70)
\Photon(30,30)(70,30){3}{4}\Photon(30,70)(70,70){3}{4}
\Text(0,80)[l]{$\bar{d}$}\Text(0,20)[l]{$s$}\Text(100,80)[r]{$e^+$}
\Text(100,20)[r]{$\mu^-$}\Text(50,20)[]{$W^-_L$}\Text(50,82)[]{$W^+_L$}
\Text(25,50)[r]{$u_i$}\Text(75,50)[l]{$n_\alpha$}
\Text(50,0)[]{\bf (a)}

\ArrowLine(150,30)(180,30)\ArrowLine(180,30)(180,70)\ArrowLine(180,70)(150,70)
\ArrowLine(220,30)(250,30)\ArrowLine(220,70)(220,30) \ArrowLine(250,70)(220,70)
\Photon(180,30)(220,30){3}{4}\Photon(180,70)(220,70){3}{4}
\Text(150,80)[l]{$\bar{d}$}\Text(150,20)[l]{$s$}\Text(250,80)[r]{$e^+$}
\Text(250,20)[r]{$\mu^-$}\Text(200,19)[]{$W_R^-$}\Text(200,82)[]{$W^+_R$}
\Text(175,50)[r]{$u_i$}\Text(225,50)[l]{$n_\alpha$}
\Text(200,0)[]{\bf (b)}

\ArrowLine(300,30)(330,30)\ArrowLine(330,30)(330,70)\ArrowLine(330,70)(300,70)
\ArrowLine(370,30)(400,30)\ArrowLine(370,70)(370,30)\ArrowLine(400,70)(370,70)
\DashArrowLine(330,30)(370,30){5}\DashArrowLine(330,70)(370,70){5}
\Text(300,80)[l]{$\bar{d}$}\Text(300,20)[l]{$s$}\Text(400,80)[r]{$e^+$}
\Text(400,20)[r]{$\mu^-$}\Text(350,19)[]{$h^-$}\Text(350,82)[]{$h^+$}
\Text(325,50)[r]{$u_i$}\Text(375,50)[l]{$n_\alpha$}
\Text(350,0)[]{\bf (c)}
\end{picture}\\[0.7cm]
\end{center}
\begin{list}{}{\labelwidth1.6cm\leftmargin2.5cm\labelsep0.4cm\itemsep0ex 
    plus0.2ex }
  
\item[{\small {\bf Fig.\ 2:}}] {\small Gauge-independent sub-set of
    Feynman graphs (group A) contributing to $K_L\to e\mu$}
\end{list}

\begin{center}
\begin{picture}(400,300)(0,0)
\SetWidth{0.8}

\ArrowLine(0,230)(30,230)\ArrowLine(30,230)(30,270)\ArrowLine(30,270)(0,270)
\ArrowLine(70,230)(100,230)\ArrowLine(70,270)(70,230) 
\ArrowLine(100,270)(70,270)  
\Photon(30,230)(70,230){3}{4}\Photon(30,270)(70,270){3}{4}
\Text(0,280)[l]{$\bar{d}$}\Text(0,220)[l]{$s$}\Text(100,280)[r]{$e^+$}
\Text(100,220)[r]{$\mu^-$}\Text(50,220)[]{$W_R^-$}\Text(50,282)[]{$W^+_L$}
\Text(25,250)[r]{$u_i$}\Text(75,250)[l]{$n_\alpha$}
\Text(50,200)[]{\bf (a)}

\ArrowLine(150,230)(180,230)\ArrowLine(180,230)(180,270)
\ArrowLine(180,270)(150,270)
\ArrowLine(220,230)(250,230)\ArrowLine(220,270)(220,230) 
\ArrowLine(250,270)(220,270)
\Photon(180,230)(220,230){3}{4}\Photon(180,270)(220,270){3}{4}
\Text(150,280)[l]{$\bar{d}$}\Text(150,220)[l]{$s$}\Text(250,280)[r]{$e^+$}
\Text(250,220)[r]{$\mu^-$}\Text(200,219)[]{$W_L^-$}\Text(200,282)[]{$W_R^+$}
\Text(175,250)[r]{$u_i$}\Text(225,250)[l]{$n_\alpha$}
\Text(200,200)[]{\bf (b)}

\ArrowLine(300,230)(330,230)\ArrowLine(330,230)(330,270)
\ArrowLine(330,270)(300,270)
\Photon(330,230)(360,250){-3}{3.5}\Photon(330,270)(360,250){3}{3.5}
\DashLine(360,250)(390,250){5}
\ArrowLine(400,270)(390,250)\ArrowLine(390,250)(400,230)
\Text(300,280)[l]{$\bar{d}$}\Text(300,220)[l]{$s$}\Text(325,250)[r]{$u_i$}
\Text(340,267)[lb]{$W_L^+$}\Text(340,231)[lt]{$W_R^-$}
\Text(375,260)[]{$\phi_2^{r,i}$}
\Text(400,280)[r]{$e^+$}\Text(400,220)[r]{$\mu^-$}
\Text(350,200)[]{\bf (c)}

\ArrowLine(0,130)(30,130)\ArrowLine(30,130)(30,170)\ArrowLine(30,170)(0,170)
\Photon(30,130)(60,150){-3}{3.5}\Photon(30,170)(60,150){3}{3.5}
\DashLine(60,150)(90,150){5}
\ArrowLine(100,170)(90,150)\ArrowLine(90,150)(100,130)
\Text(0,180)[l]{$\bar{d}$}\Text(0,120)[l]{$s$}\Text(25,150)[r]{$u_i$}
\Text(40,167)[lb]{$W_R^+$}\Text(40,131)[lt]{$W_L^-$}
\Text(75,160)[]{$\phi_2^{r,i}$}
\Text(100,180)[r]{$e^+$}\Text(100,120)[r]{$\mu^-$}
\Text(50,100)[]{\bf (d)}

\ArrowLine(150,130)(160,150)\ArrowLine(160,150)(150,170)
\ArrowLine(250,170)(220,170)\ArrowLine(220,170)(220,130)
\ArrowLine(220,130)(250,130)\DashLine(160,150)(190,150){5}
\Photon(220,170)(190,150){-3}{3.5}\Photon(220,130)(190,150){3}{3.5}
\Text(150,180)[l]{$\bar{d}$}\Text(150,120)[l]{$s$}\Text(250,180)[r]{$e^+$}
\Text(250,120)[r]{$\mu^-$}\Text(175,160)[]{$\phi_2^{r,i}$}
\Text(210,167)[rb]{$W_L^+$}\Text(210,131)[rt]{$W_R^-$}
\Text(225,150)[l]{$n_\alpha$}
\Text(200,100)[]{\bf (e)}

\ArrowLine(300,130)(310,150)\ArrowLine(310,150)(300,170)
\ArrowLine(400,170)(370,170)\ArrowLine(370,170)(370,130)
\ArrowLine(370,130)(400,130)\DashLine(310,150)(340,150){5}
\Photon(370,170)(340,150){-3}{3.5}\Photon(370,130)(340,150){3}{3.5}
\Text(300,180)[l]{$\bar{d}$}\Text(300,120)[l]{$s$}\Text(400,180)[r]{$e^+$}
\Text(400,120)[r]{$\mu^-$}\Text(325,160)[]{$\phi_2^{r,i}$}
\Text(360,167)[rb]{$W_R^+$}\Text(360,131)[rt]{$W_L^-$}
\Text(375,150)[l]{$n_\alpha$}
\Text(350,100)[]{\bf (f)}

\ArrowLine(55,30)(65,50)\ArrowLine(65,50)(55,70)
\ArrowLine(175,70)(165,50)\ArrowLine(165,50)(175,30)
\DashLine(65,50)(95,50){5}\DashLine(135,50)(165,50){5}
\PhotonArc(115,50)(20,0,360){3}{12}
\Text(55,80)[l]{$\bar{d}$}\Text(55,20)[l]{$s$}\Text(175,80)[r]{$e^+$}
\Text(175,20)[r]{$\mu^-$}
\Text(80,60)[]{$\phi_2^{r,i}$}\Text(150,60)[]{$\phi_2^{r,i}$}
\Text(115,82)[]{$W_L^+$}\Text(115,20)[]{$W_R^-$}
\Text(115,0)[]{\bf (g)}

\ArrowLine(225,30)(235,50)\ArrowLine(235,50)(225,70)
\ArrowLine(345,70)(335,50)\ArrowLine(335,50)(345,30)
\DashLine(235,50)(265,50){5}\DashLine(305,50)(335,50){5}
\PhotonArc(285,50)(20,0,360){3}{12}
\Text(225,80)[l]{$\bar{d}$}\Text(225,20)[l]{$s$}\Text(345,80)[r]{$e^+$}
\Text(345,20)[r]{$\mu^-$}
\Text(250,60)[]{$\phi_2^{r,i}$}\Text(320,60)[]{$\phi_2^{r,i}$}
\Text(285,82)[]{$W_R^+$}\Text(285,20)[]{$W_L^-$}
\Text(285,0)[]{\bf (h)}

\end{picture}\\[0.7cm]
\end{center}
\begin{list}{}{\labelwidth1.6cm\leftmargin2.5cm\labelsep0.4cm\itemsep0ex 
    plus0.2ex }
  
\item[{\small {\bf Fig.\ 3:}}] {\small Gauge-independent sub-set of
    Feynman graphs (group B) contributing to $K_L\to e\mu$}
\end{list}

There are tight experimental  constraints on the   masses of the  FCNC
scalars and the mixing angles,  which depend on the flavour  structure
of the   left-right model.  For  example, in   non-manifest left-right
models \cite{LSS}, the  experimental  limit   coming from the     mass
difference between $K_L$ and $K_S$ leads to the lower mass bound
\begin{equation}
\label{FCNCK0}
M_{\phi^r_2}\ \sim\ M_{\phi^i_2}\ \stackrel{\displaystyle >}{\sim}\ 9\
\mbox{TeV}\, .
\end{equation}
If we impose a manifest left right-symmetry on  the model, {\em i.e.},
$V^L= V^R = V$, the mass bound (\ref{FCNCK0}) is even more severe, and
the FCNC scalars must be heavier than 30 TeV \cite{MEP}.  On the other
hand, in the   manifest  left-right symmetric model,   the  tree-level
contribution to $K_L\to e\mu$ yields the following constraint:
\begin{equation}
\label{PHIbound}
M_{\phi_2^{r,i}}\ \stackrel{\displaystyle >}{\sim}\ 35
\left(\frac{V_{td}}{0.015}\right)^{1/2} 
\left(\frac{V_{ts}}{0.048}\right)^{1/2} 
\left(\frac{s^{\nu_i}_L}{0.01}\right)^{1/2}
\left(\frac{m_t}{180\mbox{ GeV}}\right)^{1/2}
\left(\frac{m_N}{10\mbox{ TeV}}\right)^{1/2}
\mbox{TeV}\, ,
\end{equation}
which  is comparable to   the one  obtained  from  the $K_LK_S$   mass
difference.

Another important bound comes  from the requirement that the  validity
of perturbative expansion be preserved. The unitarity bound derived in
this way is translated into the approximate inequality \cite{OE}
\begin{equation}
\label{FCNCuni}
M_{\phi_2^{r,i}}\ \stackrel{\displaystyle <}{\sim}\ 15\, M_R\, .
\end{equation}
The  above bound will   be considered throughout our  phenomenological
analysis in Section 5.

\begin{center}
\begin{picture}(400,300)(0,0)
\SetWidth{0.8}

\ArrowLine(0,230)(30,230)\ArrowLine(30,230)(30,270)\ArrowLine(30,270)(0,270)
\ArrowLine(70,230)(100,230)\ArrowLine(70,270)(70,230) 
\ArrowLine(100,270)(70,270)  
\DashArrowLine(30,230)(70,230){5}\Photon(30,270)(70,270){3}{4}
\Text(0,280)[l]{$\bar{d}$}\Text(0,220)[l]{$s$}\Text(100,280)[r]{$e^+$}
\Text(100,220)[r]{$\mu^-$}\Text(50,220)[]{$h^-$}\Text(50,282)[]{$W^+_L$}
\Text(25,250)[r]{$u_i$}\Text(75,250)[l]{$n_\alpha$}
\Text(50,200)[]{\bf (a)}

\ArrowLine(150,230)(180,230)\ArrowLine(180,230)(180,270)
\ArrowLine(180,270)(150,270)
\ArrowLine(220,230)(250,230)\ArrowLine(220,270)(220,230) 
\ArrowLine(250,270)(220,270)
\Photon(180,230)(220,230){3}{4}\DashArrowLine(180,270)(220,270){5}
\Text(150,280)[l]{$\bar{d}$}\Text(150,220)[l]{$s$}\Text(250,280)[r]{$e^+$}
\Text(250,220)[r]{$\mu^-$}\Text(200,219)[]{$W_L^-$}\Text(200,282)[]{$h^+$}
\Text(175,250)[r]{$u_i$}\Text(225,250)[l]{$n_\alpha$}
\Text(200,200)[]{\bf (b)}

\ArrowLine(300,230)(330,230)\ArrowLine(330,230)(330,270)
\ArrowLine(330,270)(300,270)
\DashArrowLine(330,230)(360,250){5}\Photon(330,270)(360,250){3}{3.5}
\DashLine(360,250)(390,250){5}
\ArrowLine(400,270)(390,250)\ArrowLine(390,250)(400,230)
\Text(300,280)[l]{$\bar{d}$}\Text(300,220)[l]{$s$}\Text(325,250)[r]{$u_i$}
\Text(340,267)[lb]{$W_L^+$}\Text(340,231)[lt]{$h^-$}
\Text(375,260)[]{$\phi_2^{r,i}$}
\Text(400,280)[r]{$e^+$}\Text(400,220)[r]{$\mu^-$}
\Text(350,200)[]{\bf (c)}

\ArrowLine(0,130)(30,130)\ArrowLine(30,130)(30,170)\ArrowLine(30,170)(0,170)
\Photon(30,130)(60,150){-3}{3.5}\DashArrowLine(30,170)(60,150){5}
\DashLine(60,150)(90,150){5}
\ArrowLine(100,170)(90,150)\ArrowLine(90,150)(100,130)
\Text(0,180)[l]{$\bar{d}$}\Text(0,120)[l]{$s$}\Text(25,150)[r]{$u_i$}
\Text(40,167)[lb]{$h^+$}\Text(40,131)[lt]{$W_L^-$}
\Text(75,160)[]{$\phi_2^{r,i}$}
\Text(100,180)[r]{$e^+$}\Text(100,120)[r]{$\mu^-$}
\Text(50,100)[]{\bf (d)}

\ArrowLine(150,130)(160,150)\ArrowLine(160,150)(150,170)
\ArrowLine(250,170)(220,170)\ArrowLine(220,170)(220,130)
\ArrowLine(220,130)(250,130)\DashLine(160,150)(190,150){5}
\Photon(220,170)(190,150){-3}{3.5}\DashArrowLine(190,150)(220,130){5}
\Text(150,180)[l]{$\bar{d}$}\Text(150,120)[l]{$s$}\Text(250,180)[r]{$e^+$}
\Text(250,120)[r]{$\mu^-$}\Text(175,160)[]{$\phi_2^{r,i}$}
\Text(210,167)[rb]{$W_L^+$}\Text(210,131)[rt]{$h^-$}
\Text(225,150)[l]{$n_\alpha$}
\Text(200,100)[]{\bf (e)}

\ArrowLine(300,130)(310,150)\ArrowLine(310,150)(300,170)
\ArrowLine(400,170)(370,170)\ArrowLine(370,170)(370,130)
\ArrowLine(370,130)(400,130)\DashLine(310,150)(340,150){5}
\DashArrowLine(340,150)(370,170){5}\Photon(370,130)(340,150){3}{3.5}
\Text(300,180)[l]{$\bar{d}$}\Text(300,120)[l]{$s$}\Text(400,180)[r]{$e^+$}
\Text(400,120)[r]{$\mu^-$}\Text(325,160)[]{$\phi_2^{r,i}$}
\Text(360,167)[rb]{$h^+$}\Text(360,131)[rt]{$W_L^-$}
\Text(375,150)[l]{$n_\alpha$}
\Text(350,100)[]{\bf (f)}

\ArrowLine(55,30)(65,50)\ArrowLine(65,50)(55,70)
\ArrowLine(175,70)(165,50)\ArrowLine(165,50)(175,30)
\DashLine(65,50)(95,50){5}\DashLine(135,50)(165,50){5}
\DashArrowArc(115,50)(20,180,360){5}
\PhotonArc(115,50)(20,0,180){3}{6.5}
\Text(55,80)[l]{$\bar{d}$}\Text(55,20)[l]{$s$}\Text(175,80)[r]{$e^+$}
\Text(175,20)[r]{$\mu^-$}
\Text(80,60)[]{$\phi_2^{r,i}$}\Text(150,60)[]{$\phi_2^{r,i}$}
\Text(115,82)[]{$W_L^+$}\Text(115,20)[]{$h^-$}
\Text(115,0)[]{\bf (g)}

\ArrowLine(225,30)(235,50)\ArrowLine(235,50)(225,70)
\ArrowLine(345,70)(335,50)\ArrowLine(335,50)(345,30)
\DashLine(235,50)(265,50){5}\DashLine(305,50)(335,50){5}
\DashArrowArcn(285,50)(20,180,0){5}
\PhotonArc(285,50)(20,180,360){3}{6.5}
\Text(225,80)[l]{$\bar{d}$}\Text(225,20)[l]{$s$}\Text(345,80)[r]{$e^+$}
\Text(345,20)[r]{$\mu^-$}
\Text(250,60)[]{$\phi_2^{r,i}$}\Text(320,60)[]{$\phi_2^{r,i}$}
\Text(285,82)[]{$h^+$}\Text(285,20)[]{$W_L^-$}
\Text(285,0)[]{\bf (h)}

\end{picture}\\[0.7cm]
\end{center}
\begin{list}{}{\labelwidth1.6cm\leftmargin2.5cm\labelsep0.4cm\itemsep0ex 
    plus0.2ex }
  
\item[{\small {\bf Fig.\ 4:}}] {\small Gauge-independent sub-set of
    Feynman graphs (group C) contributing to $K_L\to e\mu$}
\end{list}

If the  FCNC Higgs  scalars  are sufficiently heavy,  {\em  e.g.} much
heavier than    10  TeV   according    to  Eqs.\   (\ref{FCNCK0})  and
(\ref{PHIbound}), the  tree-level contribution  to ${\cal T}(\bar{K}^0
\to e\mu)$  may become smaller than  the present  experimental limits.
However,  the  one-loop  box diagrams  shown   in  Figs.\  2--5, being
independent of  the   heavy  neutral  Higgs  masses,  can  still  give
significant contributions to the decay  $K_L \to e\mu$. Therefore, the
impact of the loop   corrections on the phenomenological   predictions
should be studied  very carefully.  Thereby it  should be kept in mind
that the box diagrams  alone do  not form  a gauge-invariant  set, but
vertex and  Higgs vacuum polarization graphs   must be included.  This
issue  together with that  of  renormalization  will be discussed   in
detail in the next section.

\begin{center}
\begin{picture}(300,100)(0,0)
\SetWidth{0.8}
\ArrowLine(0,30)(30,30)\ArrowLine(30,30)(30,70)\ArrowLine(30,70)(0,70)
\ArrowLine(70,30)(100,30)\ArrowLine(70,70)(70,30) \ArrowLine(100,70)(70,70)
\DashArrowLine(30,30)(70,30){5}\Photon(30,70)(70,70){3}{4}
\Text(0,80)[l]{$\bar{d}$}\Text(0,20)[l]{$s$}\Text(100,80)[r]{$e^+$}
\Text(100,20)[r]{$\mu^-$}\Text(50,20)[]{$h^-$}\Text(50,82)[]{$W^+_R$}
\Text(25,50)[r]{$u_i$}\Text(75,50)[l]{$n_\alpha$}
\Text(50,0)[]{\bf (a)}

\ArrowLine(150,30)(180,30)\ArrowLine(180,30)(180,70)\ArrowLine(180,70)(150,70)
\ArrowLine(220,30)(250,30)\ArrowLine(220,70)(220,30) \ArrowLine(250,70)(220,70)
\Photon(180,30)(220,30){3}{4}\DashArrowLine(180,70)(220,70){5}
\Text(150,80)[l]{$\bar{d}$}\Text(150,20)[l]{$s$}\Text(250,80)[r]{$e^+$}
\Text(250,20)[r]{$\mu^-$}\Text(200,19)[]{$W_R^-$}\Text(200,82)[]{$h^+$}
\Text(175,50)[r]{$u_i$}\Text(225,50)[l]{$n_\alpha$}
\Text(200,0)[]{\bf (b)}

\end{picture}\\[0.7cm]
\end{center}
\begin{list}{}{\labelwidth1.6cm\leftmargin2.5cm\labelsep0.4cm\itemsep0ex 
    plus0.2ex }
  
\item[{\small {\bf Fig.\ 5:}}] {\small Gauge-independent sub-set of
    Feynman graphs (group D) contributing to $K_L\to e\mu$}
\end{list}

\setcounter{equation}{0}
\section{Gauge independence and renormalization}

In general, the  $W^\pm_R$  and $W^\pm_L$  propagators depend  on  the
gauge  used to remove  the unphysical  degrees  of freedom and on  the
parameter  chosen for fixing  such a gauge.  There  are many gauges to
carry  out   this  procedure,    such  as covariant   $R_\xi$  gauges,
non-covariant axial gauges,  {\em   etc}.  Hence, the   $W^\pm_R$  and
$W^\pm_L$  propagators may introduce gauge-fixing-parameter dependence
if  only an arbitrary    sub-set of loop   graphs is  considered.  For
example, considering box   graphs only to describe  the $K^0\bar{K}^0$
mixing in left-right models is found to be insufficient to cancel such
a       dependence    on     the    gauge-fixing    parameter    $\xi$
\cite{CSL&TI,WSH&AS,BLP,LSS}.

In the   following, we shall   closer examine the  gauge dependence of
${\cal T}(\bar{K}^0\to e^+\mu^-)$ on the  $W_L$ and $W_R$  propagators
in the covariant $R_\xi$ gauge. The $R_\xi$ gauge may be viewed as the
most practicable class of  gauges endowed with the manifest properties
of Lorentz invariance  and renormalizability.  In particular, we  wish
to identify  all those diagrams, which, together  with the box graphs,
form  a minimal gauge-independent set.   To this  end, we have divided
all the radiative corrections into four  groups A--D, depending on the
way that  the $W_R$ and  $W_L$ propagators enter  in the  loop.  These
assignments are represented in Figs.\ 2--5.   It is then not difficult
to  verify that  the groups  of box   graphs  A and  D  are separately
independent of the gauge-fixing parameter $\xi$.  In the other classes
of graphs, {\em i.e.}, B and C,  however, one must consider additional
diagrams  containing FCNC Higgs   scalars  and all relevant  tadpoles,
which are not shown in Figs.\ 3 and 4.  Since the self-energy diagrams
in groups B  and  C are UV  infinite,   we will simplify our   task by
proving gauge independence after   renormalization.  This is indeed  a
rather economical way to check $\xi$ independence, since a vast number
of tadpoles,   being momentum  independent,  will  drop out,  when all
relevant   renormalization   subtractions are   performed.    For that
purpose, we shall adopt the   on-shell skeleton (OSS)  renormalization
scheme, which is also discussed in Ref.\ \cite{BLP}.

The main virtue of the OSS scheme is its explicit maintenance of gauge
independence   during  the   process   of  renormalization.   The  OSS
renormalization  is based on  subtracting lower  $n$-point correlation
functions  from  the transition  amplitude  under  consideration.  The
lower   $n$-point  correlation  functions,   which  normally represent
self-energy and vertex graphs, are  formally gauge independent,  since
they are evaluated at on-shell external momenta.  More explicitly, the
OSS scheme may     be described as  follows.    Whenever  we have   to
renormalize  a    vertex   sub-graph,  {\em    e.g.},  the    one-loop
$\phi^r_2\mu^-e^+$  coupling $\Gamma_{\phi_2^r} (p^2)$ shown in Figs.\
4(c) and 4(d), we have just to make the subtraction
\begin{equation}
\label{R1}
\Gamma^{R1}_{\phi^r_2}(p^2)\ =\ \Gamma_{\phi^r_2}(p^2)
\, -\,\Gamma_{\phi^r_2}(M^2_{\phi^r_2})\, .
\end{equation}
The  above  operation  is called  $R1$   subtraction. By  analogy,  we
renormalize all the   self-energy  graphs  involving the  FCNC   Higgs
bosons, {\em e.g.}, $\Pi_{\phi^r_2}(p^2)$ in Figs.\  4(g) and 4(h), by
making two subtractions defined as
\begin{equation}
\label{R2}
\Pi^{R2}_{\phi^r_2}(p^2)\ =\ \Pi_{\phi^r_2}(p^2)\, -\,
\Pi_{\phi^r_2}(M^2_{\phi^r_2})\, -\,
(p^2-M^2_{\phi^r_2})\frac{d}{dp^2}\,
\Pi_{\phi^r_2}(p^2)\Big|_{p^2=M^2_{\phi^r_2}}\,  ,
\end{equation}
which  is characterized as $R2$  operation.  In our analysis, we shall
not study possible effects  due  to renormalization scheme  dependence
\cite{OS},  which    are  of    high order.       In  any  case,   the
lepton-flavour-violating decay $K_L\to  e\mu$ would not constitute the
best place to look for such effects.

In the following, we shall see how the box graphs in class C (Fig.\ 4)
are $\xi$ dependent  and how gauge  independence gets  restored if the
relevant Higgs self-energies and  vertex graphs are taken into account
in the OSS renormalization scheme.  The same procedure may be  applied
to the class  B.  The class A and  D are separately  gauge independent
and UV finite.  Obviously,  OSS  renormalization cannot apply  to  the
groups A and D, since they only consist of box diagrams.

\begin{center}
\begin{picture}(390,210)(0,0)
\ArrowLine(90,130)(100,160)\ArrowLine(100,160)(160,160)
\ArrowLine(160,160)(170,130)
\Photon(90,190)(100,160){3}{4}\DashArrowLine(170,190)(160,160){5}
\LongArrow(88,180)(91,171)
\Text(91,190)[lb]{$W_{L\mu}^+$}\Text(169,192)[rb]{$h^-$}
\Text(91,130)[lt]{$e^-$}\Text(169,130)[rt]{$\mu^-$}
\Text(130,150)[]{$n_\alpha$}
\Text(85,190)[r]{$k$}\Text(175,190)[l]{$k'$}
\Text(85,130)[r]{$p_1'$}\Text(175,130)[l]{$p_2'$}
\Text(130,170)[]{$p_1'\!+\!k$}
\Text(100,110)[]{${\bf (a)}$}
\Text(90,160)[r]{$T_\mu^{Wh}(p_1',p_2',k)=$}
\ArrowLine(290,130)(300,160)\ArrowLine(300,160)(360,160)
\ArrowLine(360,160)(370,130)
\DashArrowLine(290,190)(300,160){5}\Photon(370,190)(360,160){3}{4}
\LongArrow(372,180)(369,171)
\Text(293,192)[lb]{$h^+$}\Text(369,190)[rb]{$W_{L\mu}^-$}
\Text(291,130)[lt]{$s$}\Text(369,130)[rt]{$d$}
\Text(330,150)[]{$u_i$}
\Text(285,190)[r]{$k'$}\Text(375,190)[l]{$k$}
\Text(285,130)[r]{$p_1$}\Text(375,130)[l]{$p_2$}
\Text(330,170)[]{$p_2\!-\!k$}
\Text(300,110)[]{${\bf (b)}$}
\Text(290,160)[r]{$\tilde T_\mu^{hW}(p_1,p_2,k)=$}
\ArrowLine(90,20)(100,50)\ArrowLine(100,50)(160,50)\ArrowLine(160,50)(170,20)
\DashArrowLine(90,80)(100,50){5}\DashArrowLine(170,80)(160,50){5}
\Text(91,80)[lb]{$G_L^+$}\Text(169,82)[rb]{$h^-$}
\Text(91,20)[lt]{$e^-$}\Text(169,20)[rt]{$\mu^-$}
\Text(130,40)[]{$n_\alpha$}
\Text(85,80)[r]{$k$}\Text(175,80)[l]{$k'$}
\Text(85,20)[r]{$p_1'$}\Text(175,20)[l]{$p_2'$}
\Text(130,60)[]{$p_1'\!+\!k$}
\Text(100,0)[]{${\bf (c)}$}
\Text(90,50)[r]{$T^{Gh}(p_1',p_2',k)=$}
\ArrowLine(290,20)(300,50)\ArrowLine(300,50)(360,50)\ArrowLine(360,50)(370,20)
\DashArrowLine(290,80)(300,50){5}\DashArrowLine(370,80)(360,50){5}
\Text(293,82)[lb]{$h^+$}\Text(369,80)[rb]{$G_L^-$}
\Text(291,20)[lt]{$s$}\Text(369,20)[rt]{$d$}
\Text(330,40)[]{$u_i$}
\Text(285,80)[r]{$k'$}\Text(375,80)[l]{$k$}
\Text(285,20)[r]{$p_1$}\Text(375,20)[l]{$p_2$}
\Text(330,60)[]{$p_2\!-\!k$}
\Text(300,0)[]{${\bf (d)}$}
\Text(290,50)[r]{$\tilde T^{hG}(p_1,p_2,k)=$}
\end{picture}
\end{center}
\begin{list}{}{\labelwidth1.6cm\leftmargin2.5cm\labelsep0.4cm\itemsep0ex 
    plus0.2ex }
\item[{\small {\bf Fig.\ 6:}}] {\small Diagrammatic definitions of
    tree-level amplitudes pertinent to class C.}
\end{list}

\subsection{Gauge dependence of box diagrams}

We start examining the gauge dependence of the box diagrams in class C
shown in Fig.\ 4 in the $R_\xi$ gauges.  In  this class of gauges, the
gauge-boson  propagator,    {\em  e.g.},  of   the $W_L$    boson  may
conveniently be decomposed as follows:
\begin{equation}
  \label{Dmunu}
\Delta_{\mu\nu}(k)\ =\ U_{\mu\nu}(k)\, -\,
        \frac{k_\mu k_\nu}{M_W^2}\Delta_\xi (k)\, ,
\end{equation}
where
\begin{equation}
  \label{Umunu}
U_{\mu\nu}(k)\ =\ \frac{1}{k^2-M_W^2}
\Big(-g_{\mu\nu}+\frac{k_\mu k_\nu}{M_W^2}\Big)
\end{equation}
is the respective $W_L$-boson propagator in the unitary gauge and
\begin{equation}
  \label{Delxi}
\Delta_\xi (k)\ =\ \frac{1}{k^2-\xi M_W^2}
\end{equation}
is the $\xi$-dependent part of  the $W_L$-boson propagator.  Note that
$\Delta_\xi   (k)$ coincides  with   the propagators  of the  would-be
Goldstone  boson $G_L$ and the  ghost fields  $c_L,\ \bar{c}_L$, which
correspond   to $W_L$.   Similarly,  one  can   perform an   analogous
decomposition for $W_R$-boson propagator, which will depend on another
gauge-fixing parameter, {\em e.g.}, $\xi'$.   Such a decomposition  is
very useful to study the gauge independence  of the class B, which is,
however, conceptually very  similar to that of  the class C considered
here. Finally, the  propagator   for the  free charged  Higgs   bosons
$h^\pm$ is given by
\begin{equation}
  \label{Dh}
D_h(k)\ =\ \frac{1}{k^2-M_h^2}\ ,
\end{equation}
which is independent of  $\xi$.

To make use of the propagator  decomposition mentioned above, we write
the  amplitude   for the diagram   (a)  in Fig.\  4 as  a  sum  of two
sub-amplitudes: ${\cal  M}_{a} =   {\cal M}_{a}^W +  {\cal  M}_{a}^G$.
Omitting  integration over the   loop momentum, the sub-amplitudes are
given by
\begin{eqnarray}
  \label{MaW}
{\cal M}_{a}^W &=& -\, T_\mu^{Wh}(k)\tilde{T}_\nu^{hW}(-k)
\Delta^{\mu\nu}(k) D_h(k')\ , \\
  \label{MaG}
{\cal M}_{a}^G &=& -\, T^{Gh}(k)\tilde{T}^{hG}(-k)\Delta_\xi(k) D_h(k')\ .
\end{eqnarray}
where  $k$ and $k'$  are the loop momenta  carried by  the $W^+_L$ and
charged Higgs bosons $h^-$,   respectively.  In Eqs.\  (\ref{MaW}) and
(\ref{MaG}), $T_\mu^{Wh}$, $T_\mu^{Gh}$,    $\tilde{T}_\nu^{hW}$   and
$\tilde{T}^{hG}$  are the tree-level amplitudes  depicted  in Fig.\ 6. 
These tree-level amplitudes read:
\begin{eqnarray}
T_\mu^{Wh}(p_1',p_2',k) &=&-\frac{i}{c_\beta M_W}
\left(\frac{ig_w}{\sqrt{2}}\right)^2 
\bar{u}_\mu (p_2')\Big[c_\beta m_\mu B^R_{\mu\alpha}\mbox{P}_R+
\Big(s_\beta^2\delta_{\gamma\alpha}-C^L_{\gamma\alpha}\Big)m_\gamma
B^R_{\mu\gamma}\mbox{P}_L
\Big]\nonumber\\
& &\times
\frac{1}{\slash{p}_1'+\slash{k}-m_\alpha}
\gamma_\mu B^{L*}_{e\alpha}\mbox{P}_L v_e(p_1')\ ,\\
\tilde{T}_\mu^{hW}(p_1,p_2,k) &=& \frac{ic_\beta}{M_W}
\left(\frac{ig_w}{\sqrt{2}}\right)^2 \bar{v}_d(p_2)\gamma_\mu V^{L*}_{id}
\frac{1}{\slash{p}_2-\slash{k}-m_i}
\Big(m_i\mbox{P}_R -m_s\mbox{P}_L\Big)V^R_{is}u_s(p_1),\qquad\\
T^{Gh}(p_1',p_2',k) &=&-\frac{i}{c_\beta M_W^2}
\left(\frac{ig_w}{\sqrt{2}}\right)^2 
\bar{u}_\mu (p_2')\Big[c_\beta m_\mu B^R_{\mu\alpha}\mbox{P}_R+
\Big(s_\beta^2\delta_{\gamma\alpha}-C^L_{\gamma\alpha}\Big)m_\gamma
B^R_{\mu\gamma}\mbox{P}_L
\Big]\nonumber\\
& &\times
\frac{1}{\slash{p}_1'+\slash{k}-m_\alpha}
B^{L*}_{e\alpha}\Big(m_e\mbox{P}_R-m_\alpha\mbox{P}_L\Big)v_e(p_1')
\ ,\\
\tilde{T}^{hG}(p_1,p_2,k) &=& \frac{ic_\beta}{M_W^2}
\left(\frac{ig_w}{\sqrt{2}}\right)^2 \bar{v}_d(p_2) 
\Big(m_d\mbox{P}_R-m_i\mbox{P}_L\Big)V^{L*}_{id}
\frac{1}{\slash{p}_2-\slash{k}-m_i}\qquad\nonumber\\
& &\times\Big(m_i\mbox{P}_R -m_s\mbox{P}_L\Big)V^R_{is}u_s(p_1).
\end{eqnarray}
Using the propagator decomposition of Eq.\  (\ref{Dmunu}), one can now
write
\begin{equation}
{\cal M}_{a}^W\ =\ {\cal M}^U_a +{\cal M}^\xi_a\ ,
\end{equation}
with
\begin{eqnarray}
{\cal M}_a^U &=&-T_\mu^{Wh}(k)\tilde{T}_\nu^{hW}(-k)
U^{\mu\nu}(k)D_h(k')\ ,\\
{\cal M}_a^\xi &=& \frac{k^\mu k^\nu}{M_W^2}\ 
T_\mu^{Wh}(k)\tilde{T}_\nu^{hW}(-k)\Delta_\xi(k)D_h(k')\, .
\end{eqnarray}
Here, ${\cal M}_a^U$  denotes the amplitude of box  diagram 4(a) in the
unitary gauge, which does not explicitly depend on the gauge parameter
$\xi$. In the  unitary gauge,  the  unphysical  Goldstone bosons  (and
ghosts) are absent.

The tree-level amplitudes given in  Fig.\ 6 satisfy the following Ward
identities:
\begin{eqnarray}
\frac{k^\mu}{M_W}T_\mu^{Wh}(k) &=&
-T^{Gh}-E_L\, ,\\
\frac{k^\mu}{M_W}\tilde{T}_\mu^{hW}(k) &=&
\tilde{T}^{hG}+\tilde{E}_R\, , \\
\frac{k^\mu}{M_W}T_\mu^{hW}(k) &=&
T^{hG}+E_R\, ,\\
\frac{k^\mu}{M_W}\tilde{T}_\mu^{Wh}(k) &=&
-\tilde{T}^{Gh}-\tilde{E}_L\, , 
\end{eqnarray}
where we have defined
\begin{eqnarray}
\label{EL}
E_L(p_1',p_2')&=& i\left(\frac{g_w}{\sqrt{2}}\right)^2
\frac{c_\beta}{M_W^2}\  \bar{u}_\mu (p_2')B^R_{\mu\alpha}
B^{L*}_{e\alpha}m_\alpha\mbox{P}_L v_e(p_1')\ ,\\
\label{tildER}
\tilde{E}_R(p_1,p_2) &=& i\left(\frac{g_w}{\sqrt{2}}\right)^2
\frac{c_\beta}{M_W^2}\ \bar{v}_d(p_2)V^{L*}_{id}V^R_{is}m_i
\mbox{P}_R u_s(p_1)\ ,\\
\label{ER}
E_R(p_1',p_2') &=& i\left(\frac{g_w}{\sqrt{2}}\right)^2
\frac{c_\beta}{M_W^2}\ \bar{u}_\mu (p_2')B^L_{\mu\alpha}
B^{R*}_{e\alpha}m_\alpha\mbox{P}_R v_e(p_1')\ ,\\
\label{tildEL}
\tilde{E}_L(p_1,p_2) &=& i\left(\frac{g_w}{\sqrt{2}}\right)^2
\frac{c_\beta}{M_W^2}\  \bar{v}_d(p_2)V^{R*}_{id}V^L_{is}m_i
\mbox{P}_R u_s(p_1)\ .
\end{eqnarray}
With the  help  of the  above Ward   identities, the amplitude  ${\cal
  M}_a^\xi$ can then be rewritten as
\begin{equation}
{\cal M}_a^\xi\ =\ -{\cal M}_a^G+B_V^a+B_S^a\ ,
\end{equation}
where
\begin{eqnarray}
\label{Bas}
B_V^a &=& \Delta_\xi(k)D_h(k') \Big( T^{Gh}\tilde{E}_R
          +E_L\tilde{T}^{hG}\Big)\ ,\label{Bav}\\
B_S^a &=&\ \Delta_\xi(k)D_h(k') E_L\tilde{E}_R\ .
\end{eqnarray}
Following a similar procedure for the diagram 4(b), we obtain
\begin{equation}
{\cal M}_b^\xi\ =\ -{\cal M}_b^{G}+B_V^b+B_S^b\ .
\end{equation}
The final expression for the box graphs 4(a) and 4(b) takes the form
\begin{eqnarray}
\label{MNbox}
{\cal M}_{\mbox{\scriptsize box}} &\equiv& {\cal M}_a\, +\, {\cal M}_b
\nonumber\\
&=& {\cal M}_a^U + {\cal M}_b^U +{\cal M}_a^\xi+{\cal M}_b^\xi+{\cal M}_a^{G}
+{\cal M}_b^{G}\nonumber\\
&=& {\cal M}_{\mbox{\scriptsize box}}^U\, +\, B_V\, +\, B_S\, ,
\end{eqnarray}
where ${\cal M}_{\mbox{\scriptsize box}}^U={\cal M}_a^U+{\cal  M}_b^U$
is  the contribution  of  the box   diagrams evaluated  in the unitary
gauge, and
\begin{eqnarray}
\label{Bv}
B_V &=& \Delta_\xi(k)D_h(k') \Big( T^{Gh}\tilde{E}_R
+E_L\tilde{T}^{hG}+ T^{hG}\tilde{E}_L +E_R\tilde{T}^{Gh}\Big)\ ,\\
\label{Bs}
B_S &=& \Delta_\xi(k)D_h(k') \Big( E_L\tilde{E}_R+E_R\tilde{E}_L\Big)\ .
\end{eqnarray}
The  presence of the terms $B_V$  and $B_S$, which are proportional to
the gauge-dependent propagator  $\Delta_\xi(k)$, implies  that the box
diagrams  in class  C do  not  form  a  gauge-invariant set, and  that
additional vertex and self-energy graphs must be considered.

\subsection{OSS renormalization of vertex diagrams}

We now consider the  contribution of the  vertex diagrams 4(c)--(f) in
the OSS renormalization scheme.  After performing the $R1$ subtraction
given  in Eq.\ (\ref{R1}), we  obtain the following matrix element for
the vertex graphs:
\begin{eqnarray}
{\cal M}_{\mbox{\scriptsize vertex}}^{R1}&=&\sum_{\varphi=\phi^{r,i}}
\Big[\Gamma^0_\varphi(p_1',p_2')\frac{i}{p^2-M_\varphi^2}
\tilde{\Gamma}_\varphi^{R1}(p^2) +\Gamma_{\varphi}^{R1}(p^2)
\frac{i}{p^2-M_\varphi^2} \tilde{\Gamma}^0_\varphi(p_1,p_2)\Big]\ ,
\end{eqnarray}
where we have defined
\begin{eqnarray}
\Gamma^0_{\phi_2^r}(p_1',p_2')&=&-\, \frac{M_W}{g_wc_\beta}E_S(p_1',p_2');
\qquad
\Gamma^0_{\phi_2^i}(p_1',p_2')\ =\ 
-\, \frac{iM_W}{g_wc_\beta}E_P(p_1',p_2')\,,\\
\tilde{\Gamma}^0_{\phi_2^r}(p_1,p_2)&=&-\, \frac{M_W}{g_wc_\beta}
\tilde{E}_S(p_1,p_2);\qquad 
\tilde{\Gamma}^0_{\phi_2^i}(p_1,p_2)\ =\ -\, \frac{iM_W}{g_wc_\beta}
\tilde{E}_P(p_1,p_2)\, ,
\end{eqnarray}
with
\begin{eqnarray}
E_{S,P}(p_1',p_2') &=& E_R(p_1',p_2')\pm E_L(p_1',p_2')\ ,\\
\tilde E_{S,P}(p_1,p_2) &=& \tilde E_R(p_1,p_2)\pm \tilde E_L(p_1,p_2)\ .
\end{eqnarray}
The functions $E_L$, $\tilde{E}_R$, $E_R$, and $\tilde{E}_L$ have
previously been defined in Eqs.\ (\ref{EL})--(\ref{tildEL}), respectively.

We now proceed by decomposing ${\cal M}_{\mbox{\scriptsize vertex}}^{R1}$ 
in terms of an $R1$-subtracted amplitude in the  unitary gauge and the
remainder, {\em i.e.},
\begin{equation}
\label{MvertR1}
{\cal M}_{\mbox{\scriptsize vertex}}^{R1}\ =\   
{\cal M}_{\mbox{\scriptsize   vertex}}^{U,R1}\, +\, V_S\, +\, V_B\, .
\end{equation}
In  analogy with    Eqs.\ (\ref{MaW}) and   (\ref{MaG}),  the one-loop
vertices  involving the  virtual states  $W_L$ and  $G_L$ may  also be
separated into  two  terms: $\Gamma_{\varphi}   = \Gamma_{\varphi}^W +
\Gamma_{\varphi}^G$ and for $\tilde{\Gamma}_{\varphi}$ likewise.
Employing the identity
\begin{equation}
\Gamma_{\varphi}^{G,R1}(p^2)\ =\ \frac{p^2-M_\varphi^2}{M_h^2-M_\varphi^2}
\Gamma_{\varphi}^G(p^2)\, -\, \Big[ \frac{p^2-M_h^2}{M_h^2-M_\varphi^2}
\Gamma_{\varphi}^G(p^2)\Big]^{R1}\, ,
\end{equation}
we can cast the different terms on the RHS of Eq.\ (\ref{MvertR1}) into
the form
\begin{eqnarray}
\label{MUR1VS}
{\cal M}_{\mbox{\scriptsize vertex}}^{U,R1}+V_S &=&
\sum_{\varphi=\phi_2^{r,i}} \Big\{ \Gamma^0_\varphi\frac{i}{M_h^2-M_\varphi^2}
\Big[ \tilde{\Gamma}_{\varphi}^W(p^2) -
\frac{p^2-M_h^2}{M_h^2-M_\varphi^2}\tilde{\Gamma}_{\varphi}^G(p^2)\Big]^{R1}
\nonumber\\
 & &+\Big[ \Gamma_{\varphi}^W(p^2)- 
\frac{p^2-M_h^2}{M_h^2-M_\varphi^2}\Gamma_{\varphi}^G(p^2)
\Big]^{R1}\frac{i}{p^2-M_\varphi^2}
\tilde{\Gamma}^0_\varphi\Big\}\ ,\\
\label{VB}
V_B &=& \sum_{\varphi=\phi_2^{r,i}} \Big[
\Gamma^0_\varphi\frac{i}{M_h^2-M_\varphi^2}
\tilde{\Gamma}_{\varphi}^G(p^2) + \Gamma_{\varphi}^G(p^2)
\frac{i}{M_h^2-M_\varphi^2} \tilde{\Gamma}^0_\varphi\Big]\ .
\end{eqnarray}

Using  the Feynman rules   given in Appendix  A,  we can calculate the
analytic expressions for the one-loop vertices. These are given by
\begin{eqnarray}
\Gamma_{\phi_2^r}^G(p^2)\!\!&=&\!\!
-\frac{ig_wc_\beta}{2M_W}(M_h^2-M_{\phi_2^r}^2)\Delta_\xi (k)D_h(k')
\Big(
T^{hG}+T^{Gh}
\Big) ,\qquad\label{vfirst}\\
\Gamma_{\phi_2^i}^G(p^2)\!\!&=&\!\!
\frac{g_wc_\beta}{2M_W}(M_h^2-M_{\phi_2^i}^2)\Delta_\xi (k)D_h(k')
\Big(
T^{hG}-T^{Gh}
\Big),\qquad\\
\tilde{\Gamma}_{\phi_2^r}^G(p^2)\!\!&=&\!\!
-\frac{ig_wc_\beta}{2M_W}(M_h^2-M_{\phi_2^r}^2)\Delta_\xi (k)D_h(k')
\Big(
\tilde T^{hG}+\tilde T^{Gh}
\Big),\qquad\\
\tilde{\Gamma}_{\phi_2^i}^G(p^2)\!\!&=&\!\!
\frac{g_wc_\beta}{2M_W}(M_h^2-M_{\phi_2^i}^2)\Delta_\xi (k)D_h(k')
\Big(
\tilde T^{hG}-\tilde T^{Gh}
\Big),\qquad\\
\Gamma_{\phi_2^r}^W(p^2)\!\!&=&\!\!
\frac{ig_wc_\beta}{2}(M_h^2-M_{\phi_2^r}^2)
\Big[U^{\mu\nu}(k)-\frac{k^\mu k^\nu}{M_W^2}\Delta_\xi (k)\Big]
D_h(k')
\Big(
T^{hW}_\nu-T^{Wh}_\nu 
\Big)(k'+p)_\mu ,\,\,\qquad\\
\Gamma_{\phi_2^i}^W(p^2)\!\!\!&=&\!\!\!
-\frac{g_wc_\beta}{2}(M_h^2-M_{\phi_2^i}^2)\Big[U^{\mu\nu}(k)-
\frac{k^\mu k^\nu}{M_W^2}\Delta_\xi (k)\Big]D_h(k')
\Big(
T^{hW}_\nu\!+\!T^{Wh}_\nu
\Big)(k'+p)_\mu ,\,\,\qquad\\
\tilde{\Gamma}_{\phi_2^r}^G(p^2)\!\!&=&\!\!
\frac{ig_wc_\beta}{2}(M_h^2-M_{\phi_2^r}^2)\Big[U^{\mu\nu}(k)-
\frac{k^\mu k^\nu}{M_W^2}\Delta_\xi (k)\Big]D_h(k')
\Big(
\tilde T^{Wh}_\nu -\tilde T^{hW}_\nu
\Big)(k'+p)_\mu,\,\,\qquad\\
\tilde{\Gamma}_{\phi_2^i}^G(p^2)\!\!&=&\!\!
-\frac{g_wc_\beta}{2}(M_h^2-M_{\phi_2^i}^2)\Big[U^{\mu\nu}(k)-
\frac{k^\mu k^\nu}{M_W^2}\Delta_\xi (k)\Big]D_h(k')
\Big(
\tilde T^{hW}_\nu\!+\!
\tilde T^{Wh}_\nu
\Big)(k'+p)_\mu .\,\,\qquad\label{vlast}
\end{eqnarray}
With the help of these expressions, it is now straightforward to show
that 
\begin{equation}
V_B\ =\ -B_V\ .
\end{equation}
Evidently, in the evaluation of  the total matrix element, $V_B$  will
cancel against the  $\xi$-dependent term  $B_V$  coming  from the  box
diagrams ({\em cf.}\ (\ref{Bv})).

We will  now determine the  analytic form of the  $\xi$-dependent term
$V_S$  in  Eq.\ (\ref{MUR1VS}).   For this  purpose,   it is useful to
introduce the abbreviation
\begin{equation}
\bar{\Gamma}_{\varphi}(p^2)\ \equiv\ 
\Gamma_{\varphi}^W(p^2)-\frac{p^2-M_h^2}{M_h^2-M_\varphi^2}
\Gamma_{\varphi}^G(p^2)\ .   
\end{equation}
We can now express $\bar{\Gamma}_{\varphi^r_2}(p^2)$ as follows:
\begin{eqnarray}
\bar{\Gamma}_{\phi_2^r}(p^2) &=&
\Gamma_{\phi_2^r}^U(p^2) - \frac{ig_wc_\beta}{2M_W}
\Big( T^{hG}+T^{Gh} \Big)\Delta_\xi (k) + \frac{ig_wc_\beta}{2M_W}
E_S\Delta_\xi (k)\qquad
\nonumber\\
 & &
-\frac{ig_wc_\beta}{2M_W}(p^2-M_h^2)E_S
\Delta_\xi (k)D_h(k').
\end{eqnarray}
Here,   $\Gamma_{\phi_2^r}^U       (p^2)$    denotes    the   one-loop
$\phi^r_2\mu^-e^+$  coupling, evaluated in  the unitary  gauge.  Since
the second and the third term vanish under the $R1$ operation, as they
do not depend on $p^2$, we then obtain for the OSS renormalized vertex
function
\begin{equation}
\label{VertR1}
\bar{\Gamma}^{R1}_{\phi_2^r}(p^2)\ =\ \Gamma_{\phi_2^r}^{U,R1}(p^2)
-\frac{ig_wc_\beta}{2M_W}E_S \Big[ (p^2-M_h^2)\Delta_\xi (k)D_h(k')
\Big]^{R1}\, .
\end{equation}
The first term on the RHS of Eq.\ (\ref{VertR1}) is needed to obtain
${\cal M}_{\mbox{\scriptsize vertex}}^{U,R1}$ in Eq.\ (\ref{MUR1VS}),
whereas the remainder is a function that depends on $\xi$ explicitly.
Including both $\phi^r_2$ and $\phi^i_2$ contributions, we obtain the
$\xi$-dependent function $V_S$
\begin{eqnarray}
\label{Vs}
V_S &=& -\Big[(p^2-M_h^2)\Delta_\xi (k)D_h(k')\Big]^{R1}
\Big( E_S\tilde E_S\frac{1}{p^2-M_{\phi_2^r}^2}
-E_P\tilde E_P\frac{1}{p^2-M_{\phi_2^i}^2} \Big).
\end{eqnarray}
In the following,  we shall show  how the  $\xi$-dependent terms $V_S$
and  $B_S$, defined in  Eqs.\ (\ref{Vs}) and (\ref{Bs}), respectively,
will cancel against   corresponding $\xi$-dependent terms coming  from
the self-energy graphs, shown in Figs.\ 4(g) and 4(h).

\subsection{OSS renormalization of self-energy diagrams}

We shall now calculate the amplitude related to the self-energy graphs
(g)--(h)  in Fig.\ 4. After carrying  the  $R2$ subtraction defined in
Eq.\   (\ref{R2}), we   may  conveniently write   the self-energy-like
amplitude as
\begin{equation}
{\cal M}_{\mbox{\scriptsize self}}^{R2}\ =\ -\sum_{\varphi=\phi_2^{r,i}}
\Gamma^0_\varphi\, \frac{1}{p^2-M_\varphi^2}\, 
\Pi_\varphi^{R2}(p^2)\, \frac{1}{p^2-M_\varphi^2}\, \tilde\Gamma^0_\varphi\ .
\end{equation} 
By analogy, we also write the self-energy of the FCNC scalars as a sum
of two terms:
\begin{equation}
\label{separate}
\Pi_\varphi (p^2)\ =\ \Pi_\varphi^W (p^2)\, +\, \Pi_\varphi^G (p^2)\, ,
\end{equation}
where $\Pi_\varphi^W (p^2)$ and $\Pi_\varphi^G (p^2)$ are respectively
the graphs with $W_L$ and $G_L$ bosons in the loop. The action of $R2$
operation on the self-energy $\Pi_\varphi^G (p^2)$ gives
\begin{eqnarray}
\label{pigr2}
\Pi^{G,R2}_\varphi(p^2) &=& \Big(\, \frac{p^2-M_\varphi^2}{M_h^2-M_\varphi^2}
\, \Big)^2\Pi_\varphi^G(p^2) - 2\frac{p^2-M_\varphi^2}{(M_h^2-M_\varphi^2)^2}
\Big[(p^2-M_h^2)\Pi_\varphi^G(p^2)\Big]^{R1}\nonumber\\
& & +\, \Big[\Big(\frac{p^2-M_h^2}{M_h^2-M_\varphi^2}\Big)
\Pi_\varphi^G(p^2)\Big]^{R2}\ .
\end{eqnarray}
With the help of Eq.\ (\ref{pigr2}), the self-energy-like amplitude 
may be decomposed as follows:
\begin{equation}
  \label{msr2}
{\cal M}_{\mbox{\scriptsize self}}^{R2}\ =\ 
{\cal M}_{\mbox{\scriptsize self}}^{U,R2}\, +\, S_B\, +\, S_V\, ,
\end{equation}
where
\begin{eqnarray}
\label{MUR2}
{\cal M}_{\mbox{\scriptsize self}}^{U,R2}
&=& -\sum_{\varphi=\phi_2^{r,i}}\Gamma^0_\varphi
 \Big[\Big(\frac{p^2-M_h^2}{M_h^2-M_\varphi^2}\Big)
 \Pi_\varphi^G(p^2)+\Pi_\varphi^W(p^2)\Big]^{R2}\ 
 \tilde\Gamma^0_\varphi\, ,\\
\label{Sb}
S_B &=& -\sum_{\varphi=\phi_2^{r,i}}\Gamma^0_\varphi
 \frac{1}{(M_h^2-M_\varphi^2)^2}\Pi_\varphi^G(p^2)\ 
 \tilde\Gamma^0_\varphi\, ,\\
\label{Sv}
S_V &=& \sum_{\varphi=\phi_2^{r,i}}\Gamma^0_\varphi
 \frac{2}{(M_h^2-M_\varphi^2)(p^2-M_\varphi^2)}
 \Big[(p^2-M_h^2)\Pi_\varphi^G(p^2)\Big]^{R1}\ 
 \tilde\Gamma^0_\varphi\ .
\end{eqnarray}

We should now show that ${\cal M}_{\mbox{\scriptsize self}}^{U,R2}$ in
Eq.\ (\ref{MUR2})   is indeed  the OSS   renormalized self-energy-like
amplitude in  the unitary  gauge.    To make this  explicit,  we first
calculate the analytic   expressions for all the  scalar self-energies
$\Pi_{\varphi}^W(p^2)$ and $\Pi_{\varphi}^G(p^2)$ before applying  the
$R2$ subtraction.  Using the Feynman  rules  listed in Appendix A  and
omitting loop  integration, the  different  self-energy  contributions
read:
\begin{eqnarray}
\Pi_{\phi_2^r}^G(p^2)&=&\frac{2c_\beta^2}{M_W^2}
\Big(\frac{g_w}{2}\Big)^2(M_h^2-M_{\phi_2^2}^2)^2\Delta_\xi(k)D_h(k')
\ ,\\
\Pi_{\phi_2^i}^G(p^2)&=&\frac{2c_\beta^2}{M_W^2}
\Big(\frac{g_w}{2}\Big)^2(M_h^2-M_{\phi_2^2}^2)^2\Delta_\xi(k)D_h(k')
\ ,\\
\Pi_{\phi_2^r}^W(p^2)&=&2c_\beta^2
\Big(\frac{g_w}{2}\Big)^2
\Big[U_{\mu\nu}(k)-\frac{k_\mu k_\nu}{M_W^2}
\Delta_\xi(k)\Big]D_h(k')(p+k')^\mu (p+k')^\nu\ ,\\
\Pi_{\phi_2^i}^W(p^2)&=&\Pi_{\phi_2^r}^W(p^2)\ .
\end{eqnarray}
Taking these  last results into  account, we obtain for the expression
within the square brackets, which appears on the RHS of Eq.\ (\ref{MUR2}),
\begin{eqnarray}
\Big(\frac{p^2-M_h^2}{M_h^2-M_{\phi_2^r}^2}\Big)
\Pi_{\phi_2^r}^G(p^2)\!\!\!\!&+&\!\!\!\!\Pi_{\phi_2^r}^W(p^2)
\nonumber\\
&=&
\Pi_{\phi_2^r}^U(p^2)
-\frac{2c_\beta^2}{M_W^2}
\Big(\frac{g_w}{2}\Big)^2
\Big[2(p^2-M_h^2)-(k^{'2}-M_h^2)
\Big]\Delta_\xi (k).\qquad
\label{third}
\end{eqnarray}
The first term in Eq.\  (\ref{third}) is the $\phi_2^r$ self-energy in
the  unitary   gauge, whereas the  second  term,  being only linear in
$p^2$,  will   vanish after the $R2$    operation,  which requires two
subtractions. As a result, ${\cal M}_{\mbox{\scriptsize self}}^{U,R2}$
is indeed  the  OSS-renormalized  self-energy-like  amplitude in   the
unitary gauge.

We are then left with the gauge-dependent terms $S_B$ and $S_V$, which
can be conveniently re-expressed as follows:
\begin{eqnarray}
\label{SB2}
S_B &=& -\Delta_\xi(k)D_h(k')\Big(
E_R\tilde E_L+E_L\tilde E_R \Big)\ ,\\
\label{SV2}
S_V &=& \Big[(p^2-M_h^2)\Delta_\xi (k)D_h(k')\Big]^{R1}
\Big( E_S\tilde E_S\frac{1}{p^2-M_{\phi_2^r}^2}
-E_P\tilde E_P\frac{1}{p^2-M_{\phi_2^i}^2} \Big)\ .
\end{eqnarray}
By simple  comparison of Eqs.\ (\ref{SB2})  and (\ref{SV2}) with Eqs.\ 
(\ref{Bs}) and (\ref{Vs}), it is easy to see that
\begin{equation}
S_B\ =\ -B_S, \qquad S_V\ =\ -V_S
\end{equation}
This very last result concludes our proof of gauge-invariance of the 
diagrams in group C, presented in Fig.\ 4. 

To sum up, we  have shown that  the total amplitude  ${\cal M}$ of the
diagrams   shown  in Fig.\  4     is  $\xi$ independent in  the    OSS
renormalization  scheme. As it  should, ${\cal  M}$  equals the result
obtained in the unitary gauge, {\em i.e.},
\begin{eqnarray}
\label{final}
{\cal M} &=& {\cal M}_{\mbox{\scriptsize box}}\, +\,
{\cal M}_{\mbox{\scriptsize vertex}}^{R1}\, +\,
{\cal M}_{\mbox{\scriptsize self}}^{R2}\nonumber\\
&=& {\cal M}_{\mbox{\scriptsize box}}^U\, +\,
{\cal M}_{\mbox{\scriptsize vertex}}^{U,R1}\, +\, 
{\cal M}_{\mbox{\scriptsize self}}^{U,R2}\ .
\end{eqnarray}
In particular, we have seen how  the gauge-dependent terms originating
from the box  graphs  cancel   against  self-energy and  the    vertex
contributions after OSS renormalization.  A line of similar  steps may
be followed to show the gauge  independence of the group B. Therefore,
expressions analogous to   Eq.\ (\ref{final}) are  to be  used  in our
phenomenological considerations in Section 5.

\setcounter{equation}{0}
\section{Discussion and numerical results}

The branching ratio     $B(K_L\to e\mu)$ given  in   Eq.\ (\ref{BRKL})
depends on the reduced amplitudes ${\cal  A}$ and ${\cal B}$, which in
turn depend on many  kinematic parameters.  Since our analytic results
are shown to be  gauge independent in Section  4, we  shall henceforth
adopt  the Feynman--'t  Hooft  gauge in the  discussion that  follows.
First, we   will    qualitatively  estimate the   dominant    one-loop
contributions to the chirality enhanced amplitude ${\cal B}$. Then, we
will  compare    these estimates    with     those obtained for    the
$K^0\bar{K}^0$ system  and underline   the crucial difference  of  the
one-loop   results   for $K_L\to  e\mu$  with  those    found for  the
$K^0\bar{K}^0$ system.  In   addition, we  will  point out   the  main
improvements of our analysis, compared to  earlier studies.  Using the
complete analytic expressions  listed  in Appendix B, we  will present
numerical    predictions for   $B(K_L\to     e\mu)$  in manifest   and
non-manifest left-right models   and determine the allowed   parameter
space, when the  experimental  limit $B_{\mbox{\scriptsize  exp}}$  in
Eq.\ (\ref{Bexp}) is implemented.

We shall now   discuss  the qualitative behaviour of    the individual
contributions to the decay amplitude ${\cal T}(\bar{K}^0\to e^+\mu^-)$
in    the limit $m_{N_1},   m_{N_2}     \gg  M_R\gg M_W$.    In   this
heavy-neutrino  limit, the  reduced amplitude, ${\cal  A}_{LL}$, which
describes  the contribution from   the two $W_L$   bosons in the loop,
behaves as
\begin{equation}
\label{ALL}
{\cal A}_{LL}\ \approx\ \frac{1}{4}\, V^{L*}_{td}V^L_{ts}\, 
\frac{m^2_D}{m^2_M}\, \frac{m^2_t}{M^2_W}\,
\ln\Big(\frac{m^2_N}{m^2_t}\Big)\, .
\end{equation}
In deriving   Eq.\  (\ref{ALL}),  we   have  used the   approximation:
$s^{\nu_l}_L m_N \approx m_D$.  We see  that ${\cal A}_{LL}$ decreases
in magnitude for large $m_M$ values when $m_D$  is kept fixed, whereas
the  Dirac masses $m_D$   do not decouple  for  fixed values of $m_M$. 
Precisely the  existence  of  this non-decoupling window  due  to high
Dirac masses $m_D$ has led to enhanced predictions for lepton-flavour-
and universality-violating processes in  minimal extensions of  the SM
with right-handed neutrinos \cite{AP1,KPS,Pepe,PRD}.  For  ultra-heavy
neutrinos, the reduced amplitude ${\cal  A}_{RR}$ originating from two
$W_R$ bosons shows up a  behaviour quite analogous to ${\cal A}_{LL}$,
{\em i.e.},
\begin{equation}
\label{ARR}
{\cal A}_{RR}\ \approx\ \frac{1}{4}\, V^{R*}_{td}V^R_{ts}\, 
\frac{m^2_D}{m^2_M}\, \frac{m^2_t}{M^2_R}\, 
\ln\Big(\frac{m^2_N}{m^2_t}\Big)\, .
\end{equation}
In analogy to ${\cal A}_{LL}$,  ${\cal A}_{RR}$ vanishes when  $m_M\to
\infty$.

As has already  been  noticed  in \cite{GPS}, the   dominant  one-loop
contribution to ${\cal T}( \bar{K}^0\to e^+\mu^-)$  comes from the box
graphs with virtual  Goldstone  bosons $G^+_L$  and  $G^+_R$ in Figs.\
3(a) and 3(b). In Section 4  we have seen  however that the box graphs
are not gauge independent by themselves, and it is therefore important
to include the respective gauge-dependent complements originating from
graphs in Figs.\ 3(g) and 3(h).  If these additional contributions are
included,  we  observe  that the   gauge-independent reduced amplitude
$\eta {\cal B}$ is  still chirality enhanced  due to the  simultaneous
presence of left- and right-handed currents in the loop.  Expanding to
leading  order in the large  parameters $\beta\lambda_N = m^2_N/M^2_R$
(with $\beta = M^2_W/M^2_R$) and $\lambda_t = m^2_t/M^2_W$, we find
\begin{eqnarray}
\label{Bdom}
\eta {\cal B}(3a+3b+3g+3h) &\approx & \eta\, 
(V_{td}^{L*}V_{ts}^{R} B_{\mu N}^{R} B_{eN}^{L*} 
+V_{td}^{R*}V_{ts}^{L} B_{\mu N}^{L}
B_{eN}^{R*}) \beta (\lambda_t\lambda_N)^{1/2}\nonumber\\
&&\times\, \Big( 1\, +\,  \beta \lambda_t\ln\beta\, +\, 
\frac{\beta\lambda^2_t\ln\lambda_t}{\lambda_t -1}\, \Big)\nonumber\\
&\approx& \eta\, (V_{td}^{L*}V_{ts}^{R}\, +\, V_{td}^{R*}V_{ts}^{L})\,
\frac{m_t\, m_D}{M^2_R}\ .
\end{eqnarray}
In the last approximate  equality of Eq.\  (\ref{Bdom}), we have  used
again the fact that  $s^{\nu_l}_L m_N \approx m_D$ and  $\beta \ll 1$. 
In contrast  to ${\cal A}_{LL}$  and ${\cal A}_{RR}$, $\eta  {\cal B}$
does  not vanish   in the $m_M/v_R  \to  \infty$.   This  is a   novel
consequence of the left-right models and cannot occur in other models,
in  which gauge interactions conserve    chirality, such as   SU(2)$_L
\otimes$U(1)$_Y$ scenarios with right-handed neutrinos \cite{ZPC}.

Apart from restoration  of gauge independence,  it should  be stressed
again that the one-loop self-energies of the FCNC Higgs scalars play a
very  crucial r\^ole in the  phenomenology  of  $K_L\to e\mu$. In  the
Feynman--'t Hooft  gauge, the   dominant contribution to   the reduced
amplitude ${\cal B}$ is due the Higgs self-energies in Figs.\ 3(g) and
3(h).  Their contribution    relative  to the   box diagrams   may  be
determined from the ratio
\begin{equation}
\label{rFCNC}
r_{\mbox{\scriptsize box/FCNC}}\ \equiv\ 
\frac{{\cal B}(3a+3b)}{{\cal B}(3g+3h)}\ 
\approx\ \frac{\beta\lambda_t\ln\beta+
8\ln\beta/\lambda_N}{-1+\frac 12\ln(M_\phi^2/M_R^2)}\ .
\end{equation}
We see that the contribution of the FCNC Higgs self-energies to ${\cal
  T}(\bar{K}^0\to   e^+\mu^-)$ is considerably   enhanced  by a factor
$\sim 1/\beta$ in comparison with that of the box diagrams.  From Eq.\
(\ref{rFCNC}), we can readily  deduce that $r_{\rm box/FCNC}  \leq 1$,
unless $\beta\leq   0.01$  and $m_N\leq  1$    TeV.  However,  for the
parameter  range, in which the box  graphs  are dominant, we find that
the branching ratio is at least by one order of magnitude smaller than
the  present   experimental   limit $B_{\rm   exp}$    given in   Eq.\
(\ref{Bexp}).

In  the  manifest   left-right symmetric model    (LRSM) ({\em i.e.},\ 
$V^L=V^R=V$), the chirality enhanced amplitude $\eta {\cal B}$ in Eq.\ 
(\ref{Bdom}) can  lead to  a  combined constraint on $m_D$  and $M_R$,
when   the   experimental limit on    $K_L\to  e\mu$ is considered. In
particular, we find the inequality
\begin{equation}
\frac{m_t\, m_D}{M^2_R}\ \stackrel{\displaystyle <}{\sim}\ 
                                                4.\times\, 10^{-3}\, .
\end{equation}
If  we assume the lowest $W_R$-boson  mass allowed from $K^0\bar{K}^0$
mixing \cite{EG,BBR}, {\em i.e.}, $M_R = 1.5$  TeV, this then leads to
$m_D < 45$ GeV. This bound becomes much weaker for heavy $W_R$ bosons,
{\em e.g.}, $m_D < 0.5$ TeV for $M_R=5$ TeV.

In the manifest LRSM, the largest effect in the decay amplitude due to
the  box graphs comes from  the top quark, whereas charm-quark effects
are in general sub-dominant.  The relative behaviour  of the charm- to
top-quark contribution is given by the ratio
\begin{equation}
\label{rc/t}
r^{\mbox{\scriptsize box}}_{c/t}\ \approx\ 
\frac{m_c V^*_{cd}V_{cs}}{m_tV^*_{td}V_{ts}}\, 
\frac{4}{4+\beta\lambda_t\lambda_N}\ \approx\ 
\frac{8}{4+\beta\lambda_t\lambda_N}.
\end{equation}
${}$From  Eq.\  (\ref{rc/t}),  we  see  that   top-quark mass  effects
dominate  over the  charm-quark  effects  in  the box  diagrams,  when
$m_N\gg M_R$.  The situation is  very different for the $K^0\bar{K}^0$
mixing, in which charm-quark  yields the most dominant contribution in
the loop \cite{BLP}. The reason  for this  crucial difference is  that
the  one-loop box functions show  up  an enhanced behaviour, when  the
largest   fermionic mass in   the loop  is    much bigger  than $M_R$.
Obviously, in $K^0\bar{K}^0$ mixing, the  highest fermionic mass scale
is set up by the top quark, which is much smaller than the $W_R$-boson
mass, whereas the heavy  neutrinos can be  much heavier than $W_R$  in
the decay amplitude for $K_L\to  e\mu$.  In the  manifest LRSM, we can
estimate  the relative contribution of the  one-loop box functions for
the two cases mentioned above as follows:
\begin{equation}
\frac{{\cal A}_{\mbox{\scriptsize box}}(K^0-\bar{K}^0)}
{{\cal B}_{\mbox{\scriptsize box}}
(K_L\to e\mu )}\ \approx\ \frac{V^*_{cd}V_{cs}}{
(s^{\nu_e}_L)\, \beta\lambda^{1/2}_t\lambda_N^{1/2}}\ .
\end{equation}
This discussion shows how heavy Majorana neutrinos with masses $m_N\gg
M_R$  are phenomenologically very  significant for enhancing the decay
rate of $K_L\to e\mu$.

The decay $K_L\to e\mu$ can  also proceed via  a tree-level FCNC Higgs
exchange.   Therefore,   it is most    important to  compare  the FCNC
self-energy contribution  with that  of  the tree-level  amplitude. In
particular, we find the ratio
\begin{equation}
\label{Btree}
\frac{{\cal B}_{\rm tree}}{{\cal B}(3g+3h)}\ \approx\
\frac{4\pi M_R^2}{\alpha_w{\cal O}(1)M_\phi^2}\ \approx\ 
(190-380)\times \frac{M_R^2}{M_\phi^2}\ .
\end{equation}
It  is easy to see   that ${\cal B}  (3g+3h)$  is getting larger  than
${\cal B}_{\rm tree}$  for $M_\phi \stackrel{\displaystyle >}{\sim} 15
M_R$.   On  the other hand,  the  constraint that theory should remain
perturbative at high  energies gives  rise  to the unitarity bound  in
Eq.\ (\ref{FCNCuni}).  As    opposed  to $K^0\bar{K}^0$   mixing,  the
tree-level  Higgs contributions to  the decay $K_L\to e\mu$ are always
comparable to the one-loop effects, and hence both must be included in
the analysis.   This  constitutes another non-trivial  feature for the
decay  $K_L\to  e\mu$,  which  is  to be   examined  carefully in  the
numerical estimates.

In the numerical estimates, we consider two representative variants of
left-right models:  (i)  the manifest  or  pseudo-manifest LRSM  ({\em
  i.e.}\ $V^L=V^R=V$), in  which the Dirac  mass matrix $m_D$ is still
different  from the charged-lepton   mass matrix $M_l$,  and (ii)  the
non-manifest  LRSM, where $V^L=V$ but  $V^R\neq  V$.  Implementing the
results of Ref.\ \cite{GGMKO}, we  choose  natural relations and  mass
hierarchies among the  kinematic parameters of the  left-right models.
In particular, we set all heavy Higgs particles to be degenerate, {\em
  i.e.},  $M_{\phi^r_2} = M_{\phi^i_2}  = M_\phi$  and $M_h =  M_\phi$
({\em  cf.}\ Eq.\ (\ref{Masrel})).  Already   a small mass  difference
between the FCNC Higgs scalars  would be sufficient  to induce a large
negative radiative shift  in $R_b=\Gamma (Z\to b\bar{b})/\Gamma  (Z\to
\mbox{hadrons})$, which would be incompatible  with the existing tight
constraints on this LEP observable \cite{PRD}.

To reduce even further the number of  the many free parameters, we fix
$(s^{\nu_e}_L)^2 =  10^{-4}$, $(s^{\nu_\mu}_L)^2  =  0$ and the  angle
$\theta_R=\pi/4$ in   Eq.\ (\ref{BRmix}).  This choice  of heavy-light
neutrino mixings  is  in compliance with other  low-energy constraints
derived in Eqs.\ (\ref{mix})  and (\ref{mueconv}).  For simplicity, we
also assume that the  two heavy neutrinos,  $N_1$ and $N_2$,  have the
same mass,  {\em i.e.},  $m_{N_1}  =  m_{N_2} =  m_N$,  and  the heavy
neutrino mass $m_N$ is   varied from 100  GeV  up to the  perturbative
unitarity limit 50 TeV \cite{ZPC}, which  is compatible with the limit
of Eq.\ (\ref{mDbound}).  The input values for the  entries of the CKM
matrix  $V=V^L$  are  \cite{PDG}:  $V_{cd}=0.224$,     $V_{cs}=0.975$,
$V_{td}=0.015$  and  $V_{ts}=0.048$.   In  the  non-manifest LRSM, the
mixing  matrix  $V^R$ for    the right-handed  quarks  is  in  general
arbitrary and is only subject  to phenomenological constraints.  Here,
we  set $V^R_{cd}=V^R_{ts}=1$  and $V^R_{cs}  =  V^R_{td}  = 0$, which
leads to an enhanced charm-quark contribution  to the decay amplitude,
and in this respect, this model differs from the manifest LRSM.

Since   we are interested  in   investigating the individual tree  and
one-loop contributions  to the decay amplitude  ${\cal T}(\bar{K}^0\to
e^+\mu^-)$, our numerical  estimates for $B(K_L\to e\mu)$ are computed
from the four different squared matrix elements:
\begin{equation}
  \label{treeloop}
    \begin{array}{rl}
\mbox{I.}  & \ |{\cal  T}_{\mbox{\scriptsize tree}}|^2\quad
                                                     \mbox{(solid),}\\
\mbox{II.} & \ |{\cal  T}_{\mbox{\scriptsize tree}}|^2
             + 2\Re e({\cal  T}_{\mbox{\scriptsize tree}}
             {\cal T}^*_{\mbox{\scriptsize 1-loop}}) \quad
                                                      \mbox{(dashed),}\\
\mbox{III.}& \ |{\cal  T}_{\mbox{\scriptsize 1-loop}}|^2\quad
                                                      \mbox{(dotted),}\\
\mbox{IV.} & \ |{\cal  T}_{\mbox{\scriptsize tree}}
                    +{\cal  T}_{\mbox{\scriptsize 1-loop}}|^2\quad
                                                   \mbox{(dash-dotted).}
    \end{array}
\end{equation}
In Eq.\ (\ref{treeloop}), the type of lines  used in the plots for the
four different predictions  is specified within  the parentheses.  The
results  obtained from the  expressions I  and  II are those  that are
based on  a consistent loop expansion  of  the squared decay amplitude
$|{\cal T}(\bar{K}^0\to e^+\mu^-)|^2$.  For  comparison, we also  give
numerical estimates derived from the formulae III and IV.

In Fig.\ 7,   we display the  dependence of  $B(K_L\to e\mu )$  on the
heavy Majorana  mass $m_N$ for the  selected values of the $W_R$-boson
mass $M_R=1,\ 2,\  5$ and 10 TeV  in the manifest LRSM.   In agreement
with Eq.\  (\ref{Bdom}),   we find a  strong   dependence of $B(K_L\to
e\mu)$   on $m^2_N$.   This   non-decoupling  behaviour  of  the heavy
neutrinos enters via  the Dirac mass  terms $m_D$ of the  see-saw type
mass matrix  in Eq.\  (\ref{Mnu}) \cite{PRD}.  We   see that the  pure
one-loop contributions  to  the squared decay  amplitude (dotted line)
are always  comparable   to the respective  tree-level results  (solid
line).   In  the manifest  LRSM,   the $W_R$-boson  mass is   severely
constrained by the  experimental  limit on the $K^0\bar{K}^0$  mixing.
After  including QCD corrections, the authors   in \cite{EG} find $M_R
\stackrel{\displaystyle >}{\sim} 2.5$ TeV  in the limit of very  large
FCNC Higgs  masses.  In all plots  in Fig.\ 7, we take $M_\phi=15M_R$,
for definiteness.   As can be seen  from Fig.\ 7(c), for  example, for
sufficiently heavy Majorana neutrinos, $B(K_L\to e\mu)$ may exceed the
present experimental bound, which is indicated by an horizontal dotted
line in all plots.  Thus, for $M_R\approx 2.5$ TeV,  we find that $m_N
\stackrel{\displaystyle   <}{\sim} 10$ TeV, for  the  set of the input
parameters mentioned above.

Fig.\ 8  gives the numerical predictions  obtained in the non-manifest
LRSM. We use the same input parameters as those of  Fig.\ 7, except of
the fact that we now  consider the afore-mentioned $V^R$ matrix, which
significantly  suppresses the top-quark contribution in  the loop.  In
this scenario,  the chirality mass  enhancement due  to the heavy  top
quark is no longer  applicable and pure  loop-effects (dotted line) on
$K_L\to e\mu$  are at  most  10$\%$ in  comparison  to the  tree-level
results  for  the  experimentally  accessible  region.  In  fact,  the
tree-level effects on  $K_L\to   e\mu$ are more  significant  than the
manifest LRSM case.  The latter  leads  to tighter constraints on  the
heavy  Majorana   mass $m_N$.   It is interesting   to   note that the
non-manifest LRSM at hand  does not invalidate the experimental  bound
due   to $K^0\bar{K}^0$ mixing.  As a   result,  even relatively light
heavy  Majorana neutrinos  with masses  of  few  hundreds of GeV   can
account for possible lepton-flavour violating decays of $K_L$.

In Figs.\ 9 and 10, we examine the dependence of $B(K_L\to e\mu)$ as a
function of $M_R$ in the manifest and non-manifest LRSM, respectively.
We vary the $W_R$-boson mass from 0.5 to 20  TeV, for different values
of the heavy neutrino mass $m_N$.  The FCNC Higgs scalar mass $M_\phi$
is also varied  in linear dependence of $M_R$,  as is explicitly shown
in the  plots.  As is   expected, we observe   the generic feature  of
decoupling of a very large $W_R$ mass from the theory. In the manifest
LRSM (see Fig.\ 9), $B(K_L \to e\mu)$ is difficult to measure for $M_R
>  7$ TeV,  even if  rather heavy Majorana  neutrinos  and  FCNC Higgs
bosons are assumed.  In the case of  non-manifest LRSM, the respective
mass bound is slightly  weaker, {\em i.e.}, $M_R  > 10$ TeV, as can be
seen from Fig.\ 10.

It is now  interesting to  investigate the impact   of both the   FCNC
Higgs-scalar and $W_R$-boson masses on $B(K_L\to  e\mu)$. To this end,
we  present exclusion plots  in Figs.\ 11 and 12  for the manifest and
non-manifest     LRSM, respectively.    The     allowed  areas in  the
$M_\phi$-$M_R$  plane  are  determined from  the  inequality $B(K_L\to
e\mu) < B_{\mbox{\scriptsize  exp}}$, where $B(K_L\to e\mu)$ has  been
calculated using the squared matrix  elements in Eq.\ (\ref{treeloop})
for different  values of the heavy  neutrino  mass.  The solid, dotted
and dashed lines exclude the areas, in  which the `$m_N$' label is not
contained.   Clearly,   the  tree-level    results (solid  line)   are
independent of  the $W_R$ mass, and are  in qualitative agreement with
Eq.\  (\ref{PHIbound}).  In contrast,   the  results derived from  the
one-loop  expression  III in Eq.\  (\ref{treeloop})  (dotted line) are
insensitive to variations of $M_\Phi$.    We include the   constraints
obtained from  the squared matrix  element III, since they will become
significant  beyond one loop.  The  dashed lines  represent the limits
obtained from the   expression  II in Eq.\  (\ref{treeloop}),  whereas
results based on the  squared matrix element IV are  not shown.  To  a
good approximation, the latter may be represented  equally well by the
combined  graphical effect  of all  the  three lines mentioned  above.
From Figs.\ 11 and 12, we can see that a large region of the parameter
space of  the two typical LRSM's  can be excluded  by the experimental
limit on  the decay $K_L\to e\mu$.   For the values of the light-heavy
neutrino mixings $(s^{\nu_e}_L)^2 = 10^{-4}$, $(s^{\nu_\mu}_L)^2 = 0$,
a relatively low $W_R$ mass, {\em i.e.}, $M_R<1$ TeV, cannot naturally
accommodate   the   experimental data     from   $K_L\to  e\mu$    and
$K^0\bar{K}^0$  mixing, without recourse  to  excessive fine tuning in
the parameter space of the left-right models.

\setcounter{equation}{0}
\section{Conclusions}

We have  studied  the lepton-flavour-violating  decay $K_L   \to e\mu$
within the framework of SU(2)$_R\otimes$SU(2)$_L\otimes$U(1)$_{(B-L)}$
models. Experiments searching for this  decay mode are very important,
since they  can offer complementary limits on   the parameter space of
the left-right models together with  limits obtained from the  absence
of $\mu\to e\gamma$, $\mu-e$  conversion in  nuclei and $\mu\to  eee$,
and from the observed $K_LK_S$ mass difference.

We have seen that one-loop box diagrams are  not sufficient to provide
a gauge-invariant  result. Working in  the $R_\xi$ gauges and adopting
the  well-defined OSS renormalization,   we have shown  explicitly how
gauge  independence is restored  in  the transition amplitude,  ${\cal
  T}(\bar{K}^0\to   e^+\mu^-)$,  at one   loop, when the corresponding
self-energy  and  vertex graphs  involving  the FCNC Higgs scalars are
taken into  account.  In fact,  in the  Feynman--'t Hooft  gauge,  the
gauge-dependent complements of the  box graphs may become dominant for
a large  range of  the parameters.    This novel aspect  has  not been
addressed in detail in the existing literature before.

Using gauge-independent analytic  expressions for the one-loop  matrix
element, we have performed a systematic analysis studying the explicit
dependence of $B(K_L\to e\mu)$  on  the various kinematic  parameters.
If we keep the  enhanced light-heavy neutrino  mixing $s^{\nu_l}_L\sim
m_D/m_M$  fixed and  increase    the high  Dirac  mass terms   in  the
left-right  model, which arise from   the spontaneous breakdown of the
SU(2)$_L$  gauge symmetry, we obtain   a  quadratic dependence of  the
one-loop transition  amplitude  on   the  heavy neutrino   mass.    In
contrast, if  we hold  $m_D$  fixed below  the perturbative  unitarity
bound  but vary  the Majorana mass  scale  $m_M$,  then the transition
amplitude arising from the left-handed currents,  {\em e.g.}, from two
$W^+_L$ bosons, vanishes in  the limit $m_M\to \infty$, namely, ${\cal
  A}_{LL}\to  0$.     As is expected  on   theoretical  grounds, heavy
neutrinos being singlets under SU$(2)_L$  will eventually decouple  in
this limit.  Of course,  these are known  facts that have already been
observed in \cite{AP1}  and subsequently  discussed  in many  articles
\cite{IP,KPS,Pepe,PRD}.   Finally, our numerical estimates confirm the
non-decoupling behaviour of the  heavy neutrinos due to high SU(2)$_L$
Dirac mass terms.

In LRSM's  with  enhanced light-heavy neutrino mixing,  however, there
exists    an  additional   heavy-neutrino   enhancement   due   to the
simultaneous presence of  left- and right-handed  currents in the loop
\cite{GPS}.   These  chirality-changing   currents give  rise   to the
reduced  amplitude ${\cal B}$  in the decay amplitude (\ref{Tmatrix}).
Since $m_M$ is  not a iso-singlet mass  term under SU(2)$_R$, we  have
found   that  ${\cal  B}$  tends    to  the  non-vanishing  expression
$m_tm_D/M^2_R$ in Eq.\ (\ref{Bdom}) for $m_M\to  \infty$ and fixed VEV
$v_R$.  Clearly,   $m_M/v_R$ cannot  be arbitrarily large   but should
satisfy   the perturbative  bound   (\ref{mMbound}).  It may  be worth
emphasizing again that the  observed non-decoupling behaviour of heavy
neutrinos  in the reduced  amplitude ${\cal  B}$ is a  novelty in  the
left-right models and   has no analogue  in  the SM with  right-handed
neutrinos discussed above.    In  fact, we have   found  that the  new
non-decoupling phenomenon   can only originate from  amplitudes, which
violate    chirality  in the  loop,  thereby   allowing for  an active
interplay  between the    left-handed   mass  scale   $m_D$   and  the
right-handed one $m_M$.

Numerical  estimates have   shown  that a   significant  part of   the
parameter space of the left-right models may be constrained due to the
dominant contribution of ${\cal B}$ and  other low-energy data.  Since
these constraints depend in general on many independent parameters, we
have systematized the presentation of our results in a number of plots
in  Figs.\ 7 --  12.  In the numerical estimates   we have studied the
manifest as   well as a   typical non-manifest left-right   model.  To
exemplify further  the significance  of such an   analysis, we note in
passing that for $m_N\approx 10$ TeV, $s^{\nu_e}_L\approx 10^{-2}$ and
very  heavy FCNC  Higgs scalars, exclusion  plots  lead to $W_R$ boson
that must be  heavier  than about  3  TeV in the   manifest left-right
symmetric models ({\em cf.}\ Fig.\ 11(d)).

In conclusion,  if experiments at  DA$\Phi$NE or other future machines
could  substantiate a non-vanishing $B(K_L\to  e\mu)$  at the level of
$10^{-11}  - 10^{-12}$, this  would  be naturally accounted for within
left-right models  through  the novel non-decoupling  behaviour of the
heavy neutrinos in the chirality-changing amplitudes.  In the manifest
left-right   symmetric  models,  such  an   explanation would  require
relatively light $W_R$-boson masses with  $M_R\approx 2$ TeV and heavy
neutrinos   of   several TeV,    for    light-heavy neutrino   mixings
$s^{\nu_e}_L\approx 10^{-2}$, constraints which are in compliance with
all other low-energy data.

\bigskip\bigskip

\noindent   
{\bf  Acknowledgements.} K.S.  would like   to thank Ben Grinstein and
the UCSD physics department for their hospitality. The work of K.S. is
supported in part by the Volkswagen--Stiftung.

\newpage

\def\theequation{\Alph{section}.\arabic{equation}}
\begin{appendix}
  \setcounter{equation}{0}
\section{Feynman rules in the LRSM}

In this   appendix, we list all  the  relevant  Feynman rules obtained
within the  left-right models, which   govern the interactions  of the
gauge and Higgs   bosons  with leptons   and quarks, as  well as   the
trilinear self-couplings of the bosons. In  particular, we assume that
the SU(2)$_L$  weak   coupling   constant   $g_L$  is  equal  to   the
corresponding SU(2)$_R$ one $g_R$, {\em i.e.}, $g_L = g_R$.

With  the     assumptions and simplifications  mentioned    above, the
trilinear couplings among gauge, Higgs, and would-be Goldstone bosons,
which are only relevant  for $K_L\to e\mu$, are  given by (all momenta
flow into the vertex)
\begin{eqnarray}
W_R^{\mp\mu}(p)\phi_2^r(q)W_L^{\pm\nu}(r):\ \ \ \ \ \ \ \
&&-ig_wM_Wg_{\mu\nu}\, ,
\\
W_R^{\mp\mu}(p)\phi_2^i(q)W_L^{\pm\nu}(r):\ \ \ \ \ \ \ \
&&\pm g_wM_Wg_{\mu\nu}\, ,
\\
G_R^\pm (p)\phi_2^r(q)W_L^{\mp\mu}(r):\ \ \ \ \ \ \ \
&&\mp i\frac{g_w}{2}s_\beta (p-q)_\mu\, ,
\\
G_R^\pm (p)\phi_2^i(q)W_L^{\mp\mu}(r):\ \ \ \ \ \ \ \
&&-\frac{g_w}{2}s_\beta (p-q)_\mu\, ,
\\
G_L^\pm (p)\phi_2^r(q)W_R^{\mp\mu}(r):\ \ \ \ \ \ \ \
&&\mp i\frac{g_w}{2} (p-q)_\mu\, ,
\\
G_L^\pm (p)\phi_2^i(q)W_R^{\mp\mu}(r):\ \ \ \ \ \ \ \
&&\frac{g_w}{2}(p-q)_\mu\, ,
\\
h^\pm (p)\phi_2^r(q)W_L^{\mp\mu}(r):\ \ \ \ \ \ \ \
&&\mp \frac{g_w}{2}c_\beta (p-q)_\mu\, ,
\\
h^\pm (p)\phi_2^i(q)W_L^{\mp\mu}(r):\ \ \ \ \ \ \ \
&&-i\frac{g_w}{2}c_\beta (p-q)_\mu\, ,
\\
G_R^\pm (p)\phi_2^r(q)G_L^{\mp}(r):\ \ \ \ \ \ \ \
&&i\frac{g_w}{2M_W}s_\beta M^2_{\phi_2^r}\, ,
\\
G_R^\pm (p)\phi_2^i(q)G_L^{\mp}(r):\ \ \ \ \ \ \ \
&&\pm\frac{g_w}{2M_W}s_\beta M^2_{\phi_2^i}\, ,
\\
h^\pm (p)\phi_2^r(q)G_L^{\mp}(r):\ \ \ \ \ \ \ \
&&i\frac{g_w}{2M_W}c_\beta\Big(M_h^2-M^2_{\phi_2^r}\Big)\, ,
\\
h^\pm (p)\phi_2^i(q)G_L^{\mp}(r):\ \ \ \ \ \ \ \
&&\pm\frac{g_w}{2M_W}s_\beta\Big(M_h^2-M^2_{\phi_2^i}\Big)\, ,
\end{eqnarray}
where $s_\beta$ is defined in Eq.\ (\ref{h+GR}) and $\phi^{r,i}_2$ are
the FCNC Higgs bosons.

The   corresponding  couplings  of  the  gauge,   Higgs,  and would-be
Goldstone bosons to the charged leptons and  neutrinos can be read off
from the Lagrangians:
\begin{eqnarray}
{\cal L}^{W_R}_{l} &=& -\frac{g_w}{\sqrt{2}} W_R^{-\mu}\, B^R_{li}\ 
\bar{l}\gamma_\mu \mbox{P}_R n_i\quad +\quad \mbox{H.c.},\\
{\cal L}^{G_R^-}_{l} &=& -\frac{g_w}{\sqrt{2}M_W}\, s_\beta \, G_R^{-}\, 
B^R_{li}\ 
\bar{l}\Big[ m_l\mbox{P}_R - m_{n_i}\mbox{P}_L\Big] n_i\quad +\quad 
\mbox{H.c.},\\
{\cal L}^{h^-}_{l} &=& \frac{g_w}{\sqrt{2}M_W}\, c_\beta\, h^{-}\, 
\bar{l}\Big[ B^R_{li} m_l\mbox{P}_R - B^R_{lj}\Big(\, \delta_{ji}-
\frac{C^{R\ast}_{ji}}{c^2_\beta} \Big)\, m_{n_j}\mbox{P}_L\Big]n_i\quad 
+\quad \mbox{H.c.},\ \\
{\cal L}^{\phi_2^0}_{l} &=& -\frac{g_w}{2M_W} \phi_2^r\, 
\bar{l}_1\Big[ B^L_{l_1j}m_{n_j}B^{R\ast}_{l_2j}\mbox{P}_R + 
B^R_{l_1j}m_{n_j}B^{L\ast}_{l_2j}\mbox{P}_L\Big]l_2\nonumber\\
&&-\frac{ig_w}{2M_W} \phi_2^i\, 
\bar{l}_1\Big[ B^L_{l_1j}m_{n_j}B^{R\ast}_{l_2j}\mbox{P}_R - 
B^R_{l_1j}m_{n_j}B^{L\ast}_{l_2j}\mbox{P}_L\Big]l_2,
\end{eqnarray}
where the mixing  matrices $B^R$ and  $C^R$ are defined in Section  2,
and summation over repeated indices is implied.

Similarly, the interactions of the gauge, Higgs and would-be Goldstone
bosons with the quarks may be obtained from the Lagrangians
\begin{eqnarray}
{\cal L}^{W_R}_{q} &=& -\frac{g_w}{\sqrt{2}} W_R^{+\mu}\, V^R_{ij}\ 
\bar{u}_i\gamma_\mu \mbox{P}_R d_j\quad +\quad \mbox{H.c.},\\
{\cal L}^{G_R^-}_{q} &=& -\frac{g_w}{\sqrt{2}M_W}\, s_\beta \, G_R^{+}\, 
V^R_{ij}\ 
\bar{u}_i\Big[ m_{d_j}\mbox{P}_L - m_{u_i}\mbox{P}_R\Big] d_j\quad +\quad 
\mbox{H.c.},\\
{\cal L}^{h^-}_{q} &=& -\frac{g_w}{\sqrt{2}M_W}\, c_\beta\, h^{-}\, 
\bar{d}_j\Big[ V^{R*}_{ij} m_{d_j}\mbox{P}_R - V^{R*}_{ij} m_{u_i}
\mbox{P}_L\Big]u_i\quad 
+\quad \mbox{H.c.},\ \\
{\cal L}^{\phi_2^0}_{q} &=& -\frac{g_w}{2M_W} \phi_2^r\, 
\bar{d}_2\Big[ V^{L*}_{id_2}m_{u_i}V^{R}_{id_1}\mbox{P}_R + 
V^{R*}_{id_2}m_{u_i}V^{L}_{id_1}\mbox{P}_L\Big]d_1\nonumber\\
&&-\frac{ig_w}{2M_W} \phi_2^i\, 
\bar{d}_1\Big[ V^{L*}_{id_2}m_{u_i}V^{R}_{id_1}\mbox{P}_R - 
V^{R*}_{id_2}m_{u_i}V^{L}_{id_1}\mbox{P}_L\Big]l_2,
\end{eqnarray}
where $V^L$ and   $V^R$ are the  corresponding left  and right  mixing
matrices in the quark sector.

\setcounter{equation}{0}
\section{One-loop analytic results}

Here,   we   will present  the analytic   expressions  of all one-loop
results. To this end, we first define the following loop functions:
\begin{eqnarray}
 I(x, y)&=& x y \Big\{
-\frac 34\frac{1}{(1-x)(1- y)}
+ \Big[ \frac 14- \frac 32 \frac{1}{x-1}-\frac 34\frac{1}{(x-1)^2}
\Big] \frac{\ln x}{x- y}\nonumber \\
& & + 
\Big[ \frac 14- \frac 32 \frac{1}{ y-1}-\frac 34\frac{1}{( y-1)^2}
\Big] \frac{\ln y}{ y-x}\Big\}\, ,
\label{If}
\\
  J_1( x , y , z)&=&-\frac 14
\Big[ \frac{ x\ln x }{( x- z)^2( x- y)}
+\frac{ y\ln y }{( y- z)^2( y- x)}
+\frac{1}{( z- x)( z- y)}
\nonumber \\
 & &+
 \frac{\ln z}{( z- x)( y- z)}
\Big( \frac{ z}{ z- x}-\frac{ y}{ y- z}
\Big)\Big] \, ,
\label{J1f}
\\
J_2( x, y, z )&=&
\frac{ x\ln x}{(1- x)(1- z x)( y- x)}+
\frac{ y\ln y}{(1- y)(1- z y)( x- y)}\nonumber\\
&+&
\frac{ z\ln z}{(1- z)(1- z x)(1- z y)}\,  ,
\label{J2f}
\\
J_3( x, y, z)&=&
\frac{ x\ln x}{( x- z)( x-1)( x- y)}
+\frac{ y\ln y}{( y- z)( y-1)( y- x)}\nonumber\\
& &+\frac{ z\ln z}{( z-1)
( z- x)( z- y)}\, ,
\label{J3f}
\\
F_1( x, y, z )&=&
\frac{ x^{2}\ln x}{(1- x)(1- z x)( y- x)} +
\frac{ y^{2}\ln y}{(1- y)(1- z y)( x- y)}\nonumber\\
&+&
\frac{\ln z}{(1- z)(1- z x)(1- z y)} \, ,
\label{F1f}
\\
F_2( x, y, z)&=&-\frac 14\Big[
\frac{ x^2\ln x}{( x- z)( x-1)( x- y)}
 +\frac{ y^2\ln y}{( y- z)( y-1)( y- x)}
\nonumber\\
& &+\frac{ z^2\ln z}{( z-1)
( z- x)( z- y)}\Big] \, ,
\label{F2f}
\\
E_1( x, y, z, w)\!\!\!&\!\!=\!\!&\!\!\! w^2\Big[
\frac{ x\ln x}{( x- z)( w x-1)( x- y)}
+\frac{ y\ln y}{( y- z)( w y-1)( y- x)}
\nonumber\\
\!\!\!&\!\!\!\!&\!\!\!+\frac{ z\ln z}{(w z-1)
(z - x)( z- y)}\!+\!\frac{ w\ln w}{( w z-1)( w x-1)
( w y-1)}\Big]\, ,
\label{E1f}
\\
E_2( x, y, z, w)\!\!\!&\!\!=\!\!&\!\!\!-\frac{ w^2}{4}\Big[
\frac{ x^2\ln x}{( x- z)( w x-1)( x- y)}
+\frac{ y^2\ln y}{( y- z)( w y-1)( y- x)}
\nonumber\\
\!\!&\!\!\!\! &\!\!+\frac{ z^2\ln z}{( w z-1)
( z- x)( z- y)}\!
+\!\frac{\ln w}{( w z-1)( w x-1)( w y-1)}
\Big] \, .
\label{E2f} 
\end{eqnarray}

Equipped with the above loop functions  and the Feynman rules given in
Appendix A,  we can calculate the individual  contribution of  the box
diagrams, shown in   Figs.\ 2--5, to  the  decay amplitude of  $K_L\to
e\mu$. More explicitly, we have
\begin{eqnarray}
\label{2a}
{\cal A}(2a) &=&  V^{L*}_{id} V^L_{is}
 B^L_{\mu\alpha} B^{L*}_{e\alpha} I(\lambda_i,
\lambda_\alpha)\, ,
\\
\label{2b}
{\cal A}(2b)&=&\beta^2\,  V_{id}^{R*}V_{is}^{R}
B_{\mu\alpha}^{R}B_{e\alpha}^{R*} 
I(\beta\lambda_i,\beta\lambda_\alpha)\, ,
\\
 {\cal A}(2c)&=& V_{id}^{R*}V_{is}^{R}
B_{\mu\gamma}^{R}B_{e\delta}^{R*}\lambda_i^2
(\lambda_\gamma\lambda_\delta)^{1/2}(s_\beta^2\delta_{\gamma\alpha}-
C^L_{\gamma\alpha})
(s_\beta^2\delta_{\delta\alpha}-C^{L*}_{\delta\alpha})
J_1(\lambda_i ,\lambda_\alpha ,\lambda_h)\, ,\qquad
\label{2c}
\\
{\cal A}(5a)&=&V_{id}^{R*}V_{is}^{R}
B_{\mu\gamma}^{R}
B_{e\alpha}^{R*}\lambda_i(\lambda_\alpha\lambda_\gamma)^{1/2}
(s_\beta^2\delta_{\gamma\alpha}-
C_{\gamma\alpha}^L)\Big[ s_\beta^2 
E_2(\lambda_i,\lambda_\alpha,\lambda_h,\beta)
\nonumber\\
 & &-E_1(\lambda_i,\lambda_\alpha,\lambda_h,\beta)\Big]
\, ,
\label{5a}
\\
{\cal A}(5b)&=&V_{id}^{R*}V_{is}^{R}
B_{\mu\alpha}^{R}
B_{e\gamma}^{R*}\lambda_i(\lambda_\alpha\lambda_\gamma)^{1/2}
(s_\beta^2\delta_{\gamma\alpha}-
C_{\gamma\alpha}^{L*})\Big[ s_\beta^2 
E_2(\lambda_i,\lambda_\alpha,\lambda_h,\beta)
\nonumber\\
 & &-E_1(\lambda_i,\lambda_\alpha,\lambda_h,\beta)\Big]\, ,
\label{5b}\\
{\cal B}(3a)& = & \beta
V_{id}^{L*}V_{is}^{R}
B_{\mu \alpha}^{R}B_{e\alpha}^{L*}
(\lambda_i\lambda_\alpha)^{1/2}
\Big[ \Big( 1+\frac{\beta\lambda_i\lambda_\alpha}{4}\Big)
J_2(\lambda_i,\lambda_\alpha,\beta)
\nonumber\\
 & &- \frac{1+\beta}{4}F_1(\lambda_i,\lambda_\alpha,\beta)\Big] \, ,
\label{3a}\\
{\cal B}(3b) & = & \beta
V_{id}^{R*}V_{is}^{L}
B_{\mu \alpha}^{L}B_{e\alpha}^{R*}
(\lambda_i\lambda_\alpha)^{1/2}
\Big[ \Big( 1+\frac{\beta\lambda_i\lambda_\alpha}{4}\Big) 
J_2(\lambda_i,\lambda_\alpha,\beta)
\nonumber\\
 & &- \frac{1+\beta}{4}F_1(\lambda_i,\lambda_\alpha,\beta)\Big]\, ,
\label{3b}
\\
{\cal B}(4a) &=&V_{id}^{L*}V_{is}^{R}
B_{\mu\gamma}^{R}
B_{e\alpha}^{L*}(\lambda_i\lambda_\gamma)^{1/2}(s_\beta^2\delta_{\gamma\alpha}-
C_{\gamma\alpha}^L)\Big[ F_2(\lambda_i,\lambda_\alpha,\lambda_h)
\nonumber\\
 & &-\frac 14 
\lambda_i\lambda_\alpha
J_3(\lambda_i,\lambda_\alpha,\lambda_h)\Big]\, ,\label{4a}\\
{\cal B}(4b) &=&V_{id}^{R*}V_{is}^{L}
B_{\mu\alpha}^{L}
B_{e\gamma}^{R*}(\lambda_i\lambda_\gamma)^{1/2}(s_\beta^2\delta_{\gamma\alpha}-
C_{\gamma\alpha}^{L*})
\Big[ F_2(\lambda_i,\lambda_\alpha,\lambda_h)
\nonumber\\
 & &-\frac 14 
\lambda_i\lambda_\alpha
J_3(\lambda_i,\lambda_\alpha,\lambda_h)\Big]\, ,\label{4b}
\end{eqnarray}
where $\beta = M^2_W/M^2_R$, $\lambda_h =M^2_h/M^2_W$, and $\lambda_i$
and $\lambda_\alpha$ are defined after Eq.\ (\ref{FCNCtree}).

As has been    shown  in Section 4   however,  the  sum of the     box
contributions to  ${\cal B}$ is not  gauge  independent by itself. One
has to include  vertex and self-energy graphs,  which  are depicted in
Figs.\   3 and 4.  These  graphs  will restore gauge invariance, after
renormalization has been taken into account (see Section 4).  Since we
are interested in finding the corresponding gauge-dependent complement
part of  the box contributions to  ${\cal B}$ in the Feynman--'t Hooft
gauge, we keep only those terms that do  not vanish in the formal mass
limit $M_{\phi^{r,i}_2}\to \infty$.  In  Section 4, we have also  seen
that there will be an implicit dependence of the loop integrals on the
FCNC scalar masses,   which   enters  via the OSS    renormalization.  
Furthermore,  we take both the  FCNC scalars to be exactly degenerate,
{\em i.e.}, $M_{\phi^r_2}=  M_{\phi^i_2}=M_{\phi}$,  which  is a  good
approximation (see also Eq.\ (\ref{Masrel})).  In this  way, it is not
difficult to find from the would-be Goldstone-boson graphs
\begin{eqnarray}
  \label{3cd}
{\cal B}(3c+3d) &=& \frac 14\, 
V_{id}^{L*}V_{is}^{R} B_{\mu \alpha}^{R}B_{e\alpha}^{L*}\, 
s^2_\beta \lambda^{3/2}_i\lambda^{1/2}_\alpha\,  
C^{R1}_0(0,\lambda_\phi,0,\lambda_i,1,1/\beta )\nonumber\\
&&+\frac 14\, V_{id}^{R*}V_{is}^L B_{\mu \alpha}^L B_{e\alpha}^{R*}\, 
s^2_\beta \lambda^{3/2}_i\lambda^{1/2}_\alpha\,  
C^{R1}_0(0,\lambda_\phi,0,\lambda_i,1,1/\beta )\, ,\\
  \label{3ef}
{\cal B}(3e+3f) &=& \frac 14\, V_{id}^{L*}V_{is}^{R} 
B_{\mu \alpha}^{R}B_{e\alpha}^{L*}\, 
s^2_\beta \lambda^{1/2}_i\lambda^{3/2}_\alpha\,  
C^{R1}_0(0,\lambda_\phi,0,\lambda_\alpha,1,1/\beta )\nonumber\\
&&+\frac 14\, V_{id}^{R*}V_{is}^L B_{\mu \alpha}^L B_{e\alpha}^{R*}\, 
s^2_\beta \lambda^{1/2}_i\lambda^{3/2}_\alpha\,  
C^{R1}_0(0,\lambda_\phi,0,\lambda_\alpha,1,1/\beta )\, ,\\
  \label{3gh}
{\cal B}(3g+3h)&=& \frac{1}{16}\, 
V_{id}^{L*}V_{is}^{R} B_{\mu \alpha}^{R}B_{e\alpha}^{L*}\, 
\beta (\lambda_i\lambda_\alpha)^{1/2}\, 
B^{R2}_0(\lambda_\phi,1,1/\beta)\nonumber\\
&& +\, \frac{1}{16}\, 
V_{id}^{R*}V_{is}^{L} B_{\mu \alpha}^{L}B_{e\alpha}^{R*}\, 
\beta (\lambda_i\lambda_\alpha)^{1/2}\, 
B^{R2}_0(\lambda_\phi,1,1/\beta)\, ,\\
  \label{4cd}
{\cal B}(4c+4d) &=& \frac 14\, (1-\lambda_h/\lambda_\phi)\,  
V_{id}^{L*}V_{is}^{R} B_{\mu \alpha}^{R}B_{e\alpha}^{L*}\, 
c^2_\beta \lambda^{3/2}_i\lambda^{1/2}_\alpha\,  
C^{R1}_0(0,\lambda_\phi,0,\lambda_i,1,\lambda_h )\nonumber\\
&&+\, \frac 14 (1-\lambda_h/\lambda_\phi)\,
V_{id}^{R*}V_{is}^L B_{\mu \alpha}^L B_{e\alpha}^{R*}\, 
s^2_\beta \lambda^{3/2}_i\lambda^{1/2}_\alpha\,  
C^{R1}_0(0,\lambda_\phi,0,\lambda_i,1,\lambda_h )\, ,\qquad\\
  \label{4ef}
{\cal B}(4e+4f) &=& \frac 14\, (1-\lambda_h/\lambda_\phi)\, 
V_{id}^{L*}V_{is}^{R} 
B_{\mu \gamma}^{R}B_{e\alpha}^{L*} (s^2_\beta\delta_{\alpha\gamma}-
C^L_{\alpha\gamma})\, \lambda^{1/2}_i\lambda_\alpha \lambda^{1/2}_\gamma 
\nonumber\\  
&&\times\, C^{R1}_0 (0,\lambda_\phi,0,\lambda_\alpha,1,\lambda_h )\, +\, 
\frac 14\, (1-\lambda_h/\lambda_\phi)\, V_{id}^{R*}V_{is}^L
B_{\mu \alpha}^L B_{e\gamma}^{R*}\nonumber\\ 
&&\times\, (s^2_\beta\delta_{\alpha\gamma} - C^{L*}_{\alpha\gamma}) 
\lambda^{1/2}_i\lambda^{1/2}_\gamma \lambda_\alpha\,  
C^{R1}_0(0,\lambda_\phi,0,\lambda_\alpha,1,\lambda_h )\, ,\\
  \label{4gh}
{\cal B}(4g+4h)&=& \frac{1}{16}\, (1-\lambda_h/\lambda_\phi)^2\,
V_{id}^{L*}V_{is}^{R} B_{\mu \alpha}^{R}B_{e\alpha}^{L*}\, 
c^2_\beta 
(\lambda_i\lambda_\alpha)^{1/2}\, 
B^{R2}_0(\lambda_\phi,1,\lambda_h )\nonumber\\
&& + \frac{1}{16}\, (1-\lambda_h/\lambda_\phi)^2\,
V_{id}^{R*}V_{is}^{L} B_{\mu \alpha}^{L}B_{e\alpha}^{R*}\, 
\beta (\lambda_i\lambda_\alpha)^{1/2}\, 
B^{R2}_0(\lambda_\phi,1,\lambda_h )\, .
\end{eqnarray}
Since $\lambda_h -\lambda_\phi={\cal O}(1)$,  as has been discussed in
Section 2, the reduced amplitudes ${\cal B}(4c+4d)$, ${\cal B}(4e+4f)$
and   ${\cal     B}(4g+4h)$     behave  as    $1/\lambda_\phi$     and
$1/\lambda^2_\phi$.   Therefore,  these  contributions may  safely  be
neglected in the numerical estimates.

The OSS renormalization scheme adopted in  Section 4 gives rise to the
subtracted loop functions $C^{R1}_0$ and $B^{R2}_0$, which are defined
in the following way:
\begin{eqnarray}
\label{B0R2}  
B_0^{R2}(M^2,m^2_1,m^2_2)\!\! &=&\!\! B_0(0,m^2_1,m^2_2)\! -\! 
B_0(M^2,m^2_1,m^2_2)\! +\! M^2 B'_0(M^2,m^2_1,m^2_2) ,\qquad\\
\label{C0R1}  
C_0^{R1}(0,M^2,0,m^2_1,m^2_2,m^2_3)\!\! &=&\!\! C_0(0,0,0,m^2_1,m^2_2,m^2_3)
\! -\! C_0(0,M^2,0,m^2_1,m^2_2,m^2_3)\, .\qquad
\end{eqnarray}  
The  loop  functions   $B_0(M^2,m^2_1,m^2_2)$ and  $C_0(0,M^2,0,m^2_1,
m^2_2, m^2_3)$ are the usual  Passarino--Veltman functions defined  in
Ref.\ \cite{Velt/Pas}. Their explicit form is given by 
\begin{eqnarray}
   \label{B0}
B_0(M^2,m^2_1,m^2_2) &=& C_{\rm UV}\, -\, \ln (m_1m_2)\, +\, 2\, +\, 
\frac{1}{M^2}\, \Big[(m^2_2-m^2_1)\, \ln\frac{m_1}{m_2}\nonumber\\
&&+\, \lambda^{1/2}(M^2,m^2_1,m^2_2)\,\, {\rm cosh}^{-1}
\Big(\, \frac{m^2_1+m^2_2-M^2}{2m_1m_2}\, \Big)\, \Big],\\
    \label{C0}
C_0(0,M^2,0,m^2_1,m^2_2,m^2_3) &=& -\frac{1}{M^2}\, \Big[\,
{\rm Li}_2\Big(\frac{m^2_1-m^2_3}{m^2_1-m^2_3 +M^2\xi_+}\Big)\, -\,
{\rm Li}_2\Big(\frac{m^2_1-m^2_3+M^2}{m^2_1-m^2_3 +M^2\xi_+}\Big)\, \nonumber\\
&& +\, {\rm Li}_2\Big(\frac{m^2_1-m^2_3}{m^2_1-m^2_3 +M^2\xi_-}\Big)\, -\,
{\rm Li}_2\Big(\frac{m^2_1-m^2_3+M^2}{m^2_1-m^2_3 +M^2\xi_-}\Big)\, \nonumber\\
&&-\, {\rm Li}_2\Big(\frac{(m^2_1-m^2_2)(m^2_1-m^2_3)}{
(m^2_1-m^2_2)(m^2_1-m^2_3) +M^2 m^2_1}\Big)\nonumber\\
&& +\, {\rm Li}_2\Big(\frac{(m^2_1-m^2_2)(m^2_1-m^2_3+M^2)}{
(m^2_1-m^2_2)(m^2_1-m^2_3) +M^2 m^2_1}\Big)\, \Big]\, ,
\end{eqnarray}
where $\lambda (x,y,z) = (x-y-z)^2 - 4yz$,  ${\rm cosh}^{-1} z = \ln (
z + \sqrt{z^2 - 1} )$ and $C_{\rm  UV}$ is an  UV constant which drops
out after renormalization.  Furthermore,  in Eq.\ (\ref{C0}),  we have
defined 
\begin{equation}
\xi_\pm\ =\ \frac{1}{2 M^2}\, [M^2 - m^2_2 + m^2_3\, \pm\, 
\lambda^{1/2}(M^2,m^2_2,m^2_3)],
\end{equation}
and the dilogarithmic function
\begin{equation}
{\rm Li}_2 (x\pm i\varepsilon)\ =\ -\, \int_0^x\, dt\,
\frac{\ln|1-t|}{t}\ \pm\ i\theta(x-1)\, \pi\ln x\ .
\end{equation}

\end{appendix}

\newpage

\centerline{\Large {\bf Figures 7 -- 12}}

\begin{figure}[hb]
  \leavevmode
 \begin{center}
   \epsfxsize=16.2cm \epsffile[0 0 567 652]{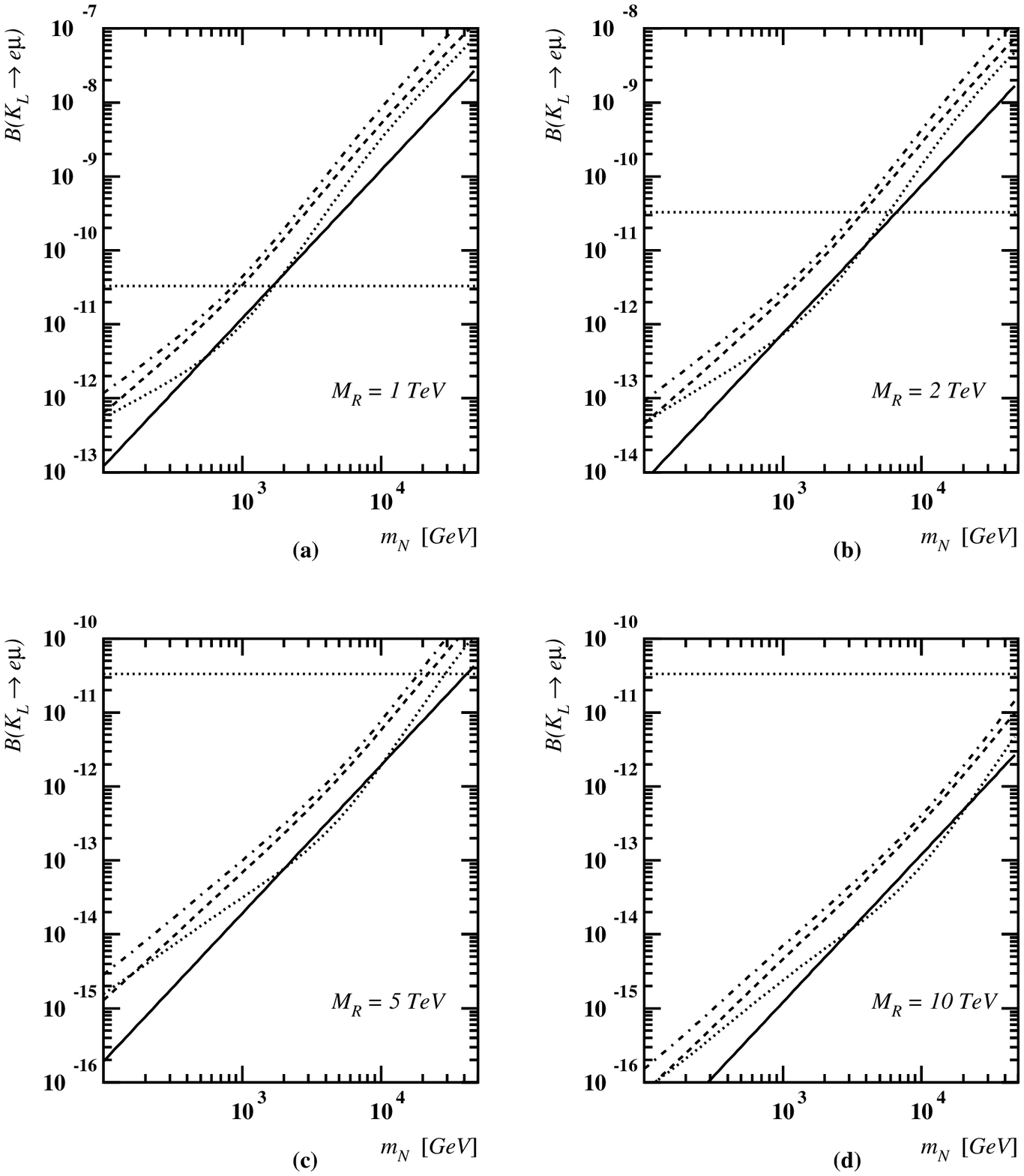}
\end{center}
\begin{list}{}{\labelwidth1.6cm\leftmargin2.5cm\labelsep0.4cm\itemsep0ex
    plus0.2ex }
\item[{\small {\bf Fig.\ 7:}}] {\small $B(K_L\to e\mu)$ versus $m_N$
    for different $M_R$ values in the manifest LRSM \\ (the meaning of
    the various lines is given in the text).}
\end{list}
\end{figure}

\newpage

\begin{figure}[hb]
  \leavevmode
 \begin{center}
   \epsfxsize=16.2cm \epsffile[0 0 567 652]{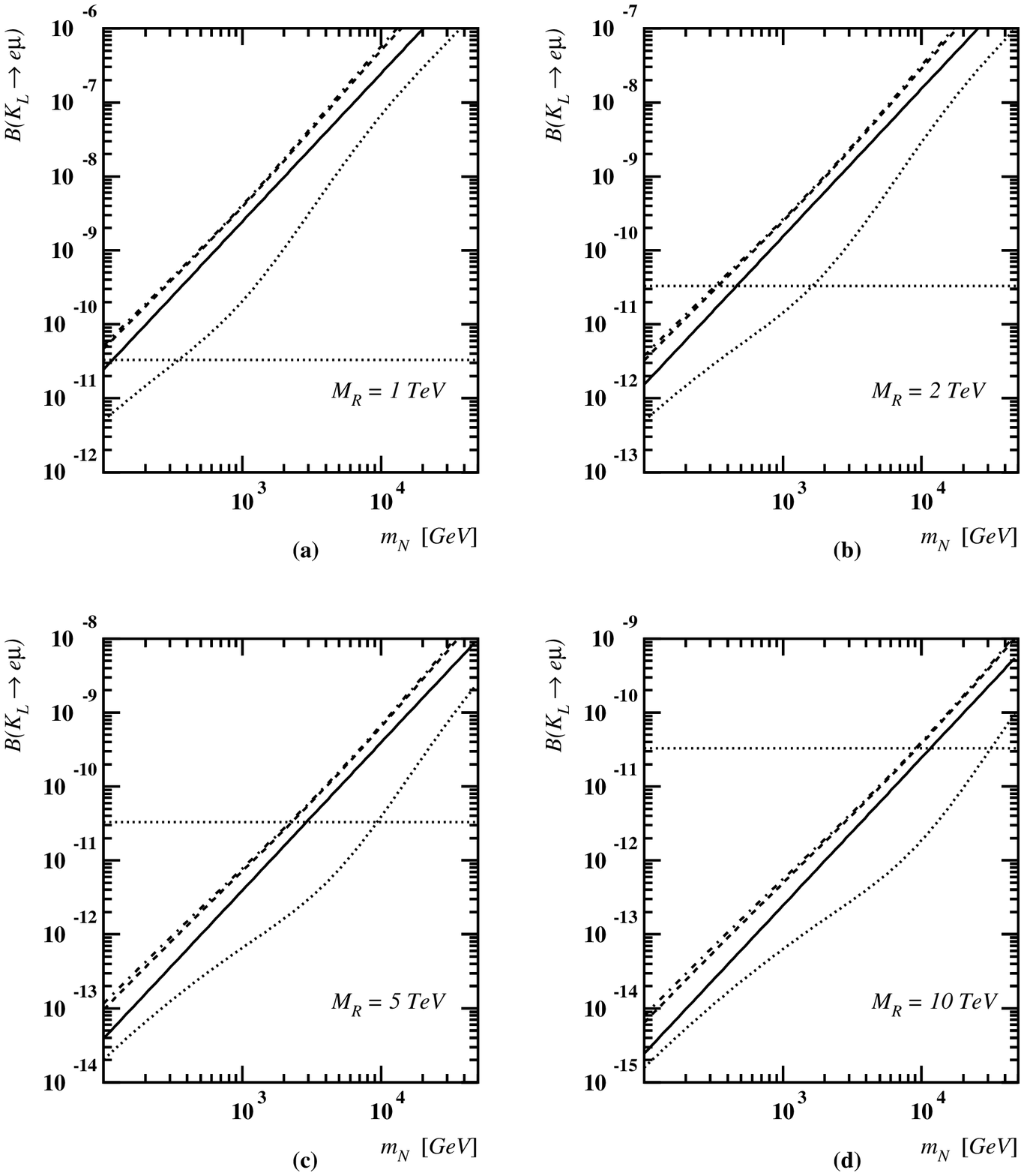}
\end{center}
\begin{list}{}{\labelwidth1.6cm\leftmargin2.5cm\labelsep0.4cm\itemsep0ex
    plus0.2ex }
\item[{\small {\bf Fig.\ 8:}}] {\small $B(K_L\to e\mu)$ versus $m_N$
    for different $M_R$ values in the non-manifest LRSM \\ (the
    meaning of the various lines is given in the text).}
\end{list}
\end{figure}

\newpage

\begin{figure}[hb]
  \leavevmode
 \begin{center}
   \epsfxsize=16.2cm \epsffile[0 0 567 652]{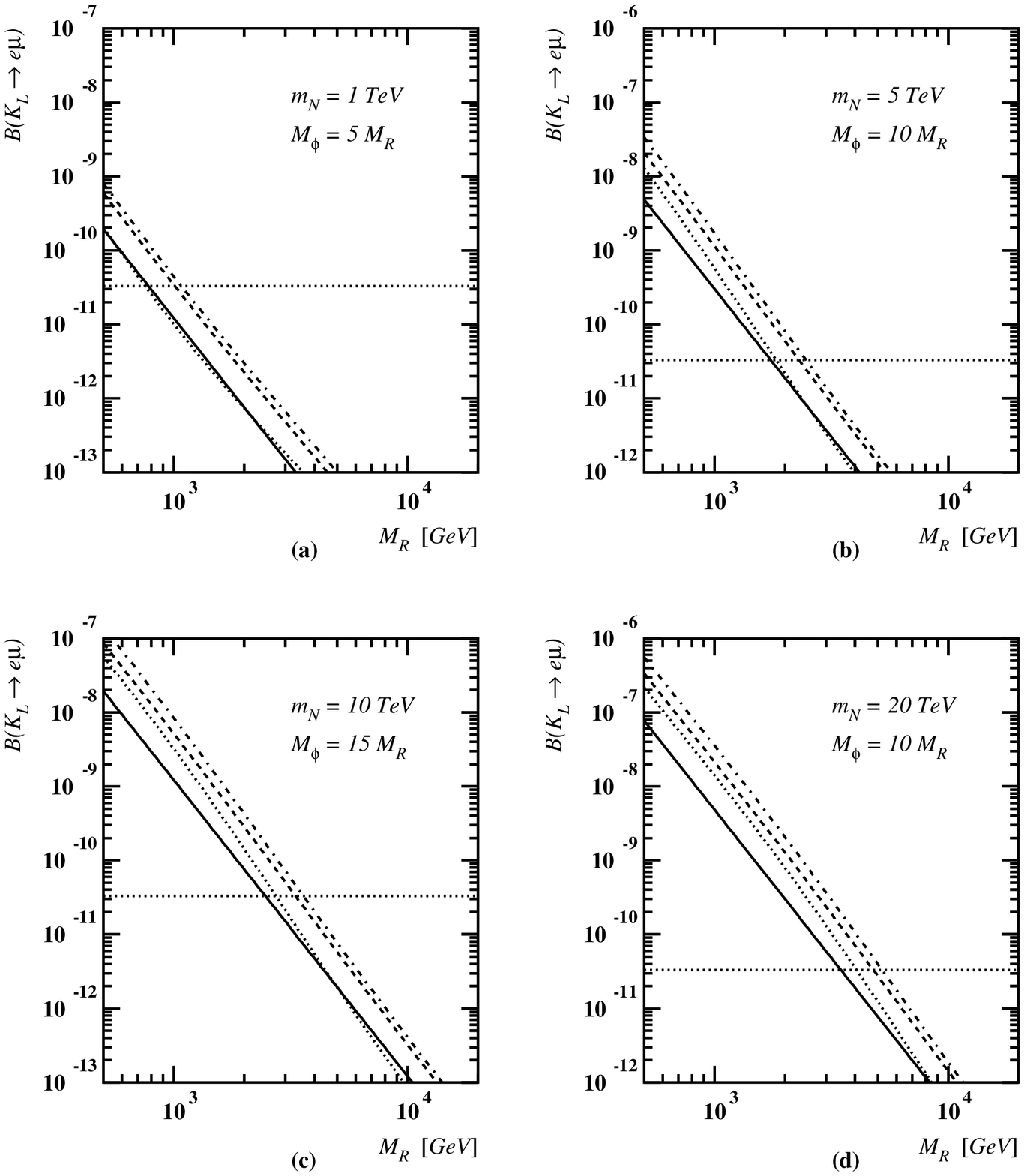}
\end{center}
\begin{list}{}{\labelwidth1.6cm\leftmargin2.5cm\labelsep0.4cm\itemsep0ex
    plus0.2ex }
\item[{\small {\bf Fig.\ 9:}}] {\small $B(K_L\to e\mu)$ versus $M_R$
    for different $m_N$ values in the manifest LRSM \\ (the meaning of
    the various lines is given in the text).}
\end{list}
\end{figure}

\newpage

\begin{figure}[hb]
  \leavevmode
 \begin{center}
   \epsfxsize=16.2cm \epsffile[0 0 567 652]{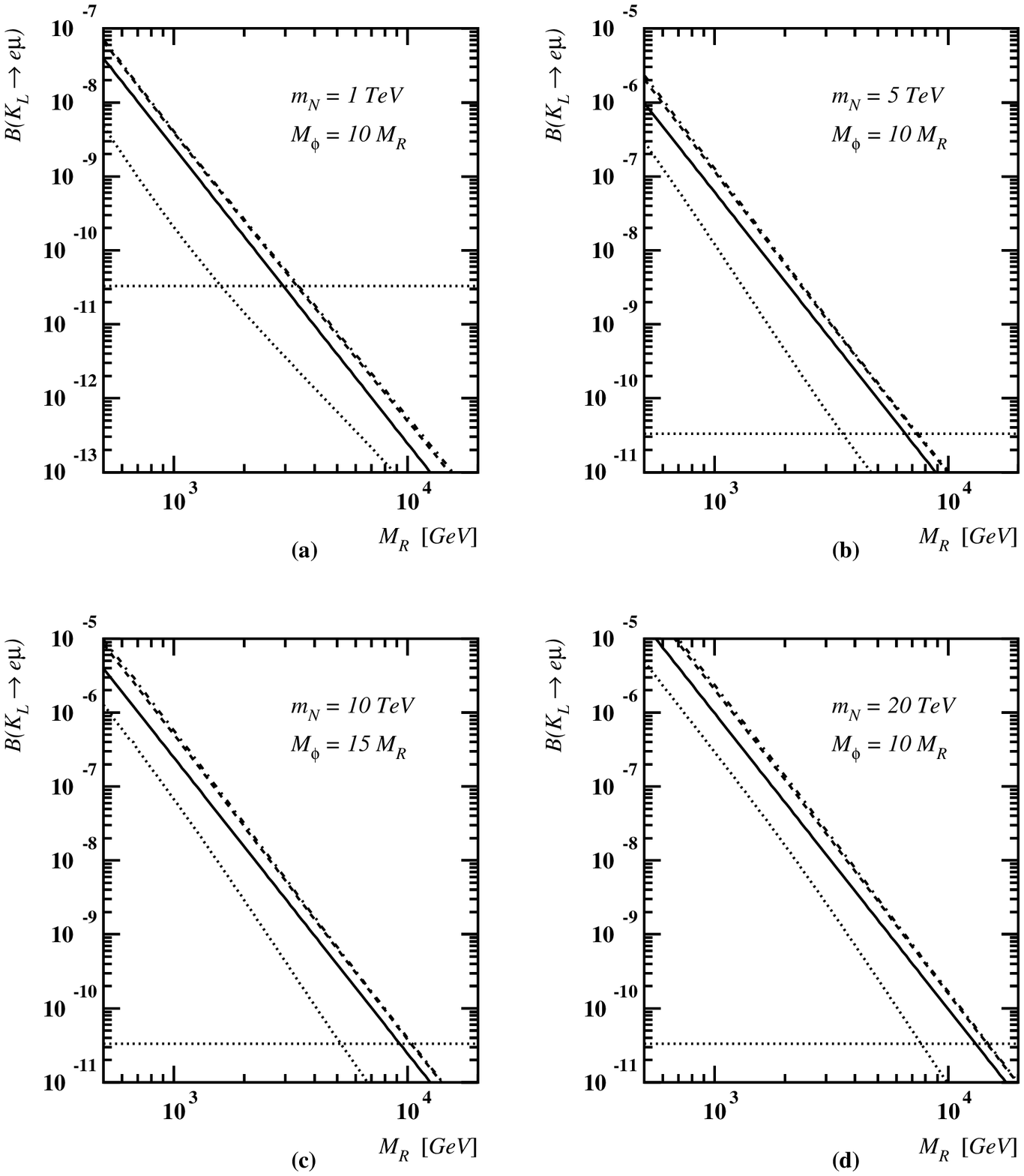}
\end{center}
\begin{list}{}{\labelwidth1.6cm\leftmargin2.5cm\labelsep0.4cm\itemsep0ex
    plus0.2ex }
\item[{\small {\bf Fig.\ 10:}}] {\small $B(K_L\to e\mu)$ versus $M_R$
    for different $m_N$ values in the non-manifest LRSM \\ (the
    meaning of the various lines is given in the text).}
\end{list}
\end{figure}

\newpage

\begin{figure}[hb]
  \leavevmode
 \begin{center}
   \epsfxsize=16.2cm \epsffile[0 0 567 652]{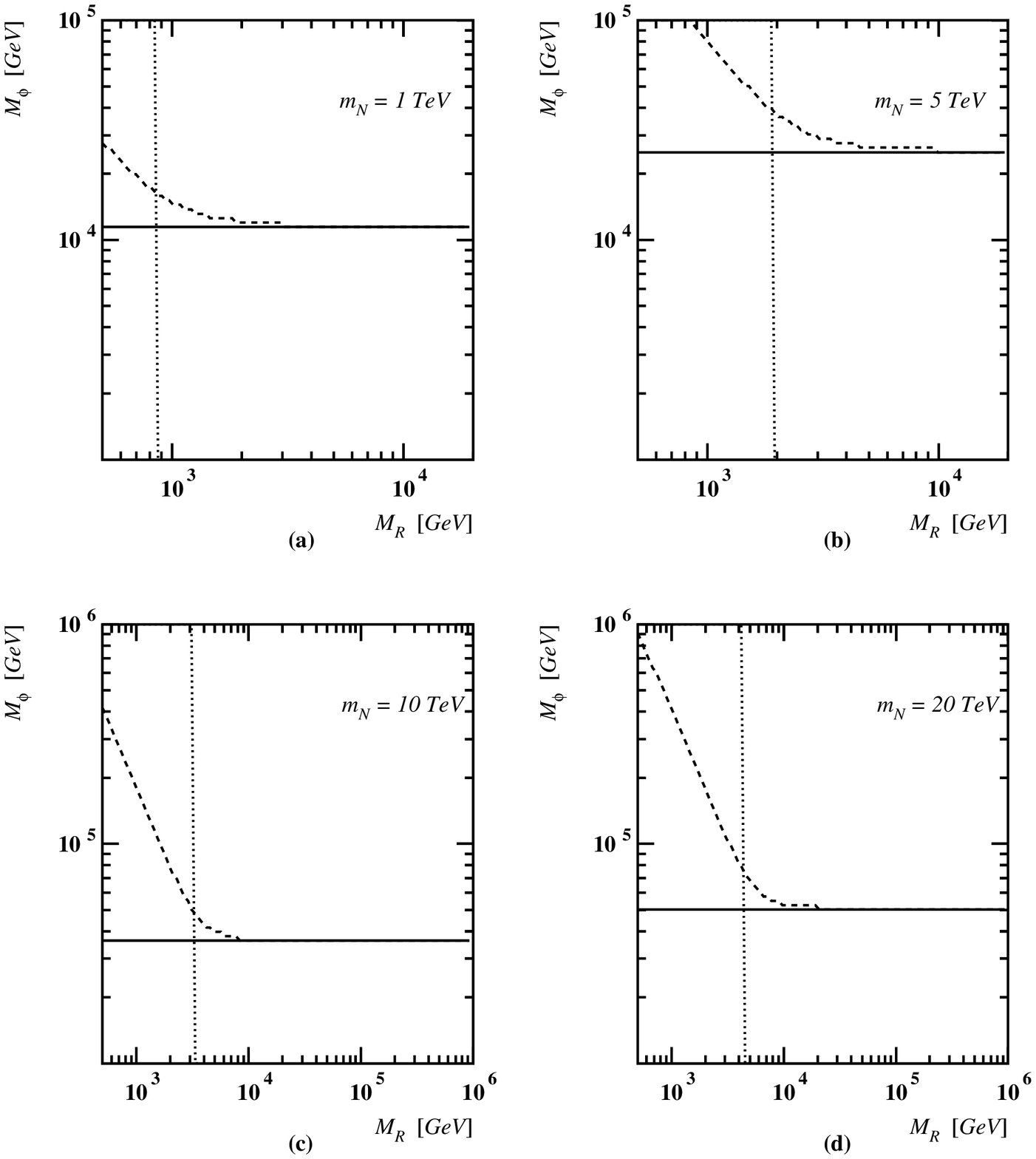}
\end{center}
\begin{list}{}{\labelwidth1.6cm\leftmargin2.5cm\labelsep0.4cm\itemsep0ex
    plus0.2ex }
\item[{\small {\bf Fig.\ 11:}}] {\small Exclusion plots for the
    parameters $M_\phi,\ M_R$, obtained from
    the constraint \\
    $B(K_L\to e\mu)<3.3\times 10^{-11}$ in the manifest LRSM
    (the meaning of the \\
    various lines is given in the text).}
\end{list}
\end{figure}

\newpage

\begin{figure}[hb]
  \leavevmode
 \begin{center}
   \epsfxsize=16.2cm \epsffile[0 0 567 652]{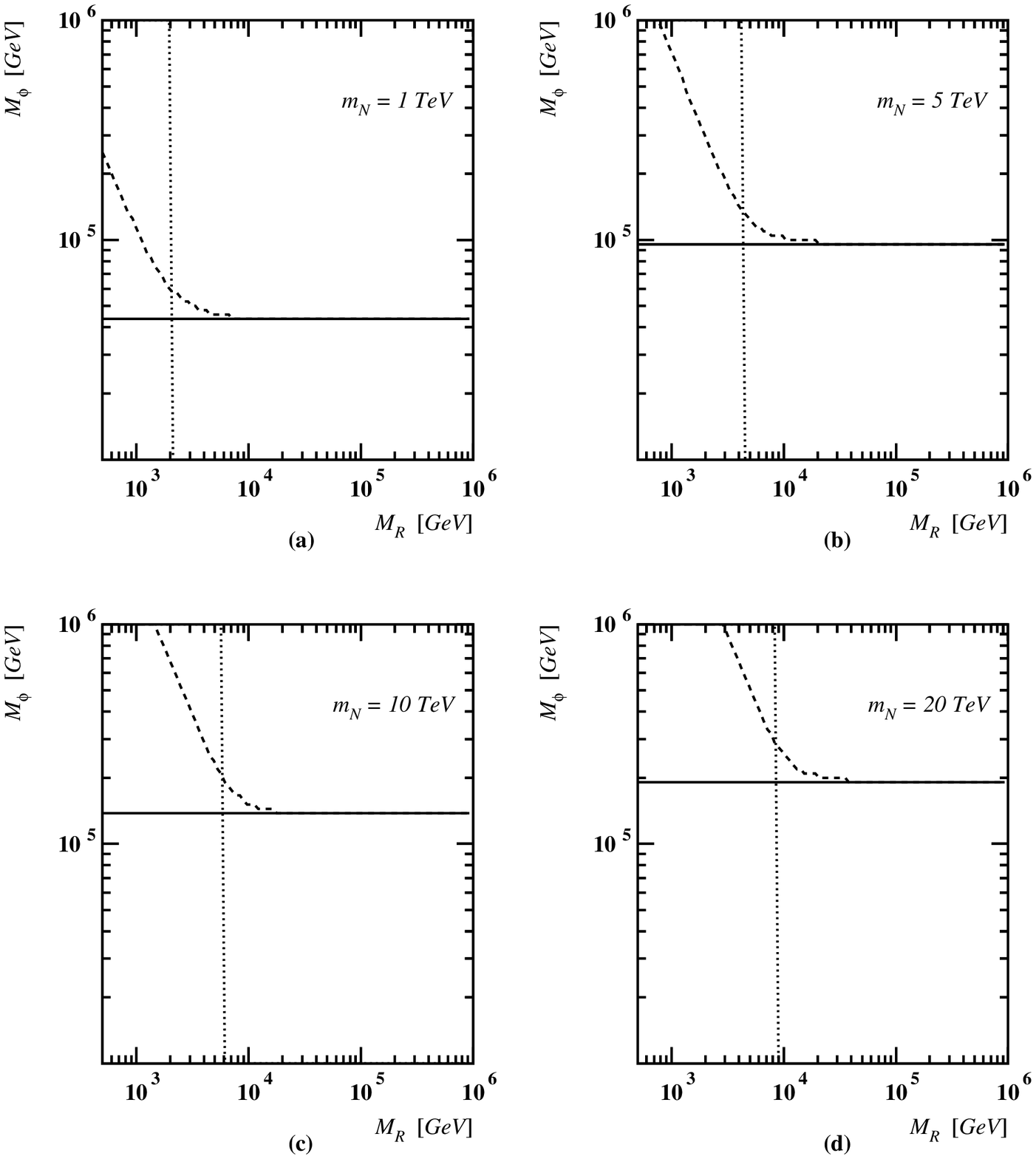}
\end{center}
\begin{list}{}{\labelwidth1.6cm\leftmargin2.5cm\labelsep0.4cm\itemsep0ex
    plus0.2ex }
\item[{\small {\bf Fig.\ 12:}}] {\small Exclusion plots for the
    parameters $M_\phi,\ M_R$, obtained from
    the constraint \\
    $B(K_L\to e\mu)<3.3\times 10^{-11}$ in the
    non-manifest LRSM    (the meaning of the \\
    various lines is given in the text).}
\end{list}
\end{figure}

\end{document}